\documentclass[emulateapj,twocolumn]{aastex63}
\usepackage{graphicx}
\usepackage{float}
\usepackage{amsmath}
\usepackage{subfigure}
\usepackage{natbib}
\usepackage{array}
\usepackage{gensymb}
\usepackage{float}
\restylefloat{table}
\newcommand{\head}[2]{\multicolumn{1}{>{\centering\arraybackslash}p{#1}}{\textbf{#2}}}
\usepackage{booktabs}
\usepackage{xcolor}

\begin{document}


\title{A large-scale kinematic study of molecular gas in high-z cluster galaxies: Evidence for high levels of kinematic asymmetry}
\author{Cramer, W. J.}
\affil{School of Earth and Space Exploration, Arizona State University, Tempe, AZ 85287, USA}
\author{Noble, A. G.}
\affil{School of Earth and Space Exploration, Arizona State University, Tempe, AZ 85287, USA}
\affil{Beus Center for Cosmic Foundations, Arizona State University, Tempe, AZ 85287-1404, USA}
\author{Massingill, K.}
\affil{School of Earth and Space Exploration, Arizona State University, Tempe, AZ 85287, USA}
\author{Cairns, J.}
\affil{Blackett Lab, Imperial College London, Prince Consort Road, London SW7 2AZ, UK}
\author{Clements, D. L.}
\affil{Blackett Lab, Imperial College London, Prince Consort Road, London SW7 2AZ, UK}
\author{Cooper, M. C.}
\affil{Department of Physics \& Astronomy, University of California, Irvine, 4129 Reines Hall, Irvine, CA 92697, USA}
\author{Demarco, R.}
\affil{Departamento de Astronom{\'i}a, Facultad de Ciencias F{\'i}sicas y Matem{\'a}ticas, Universidad de Concepci{\'o}n, Concepci{\'o}n, Chile}
\author{Matharu, J.}
\affil{Department of Physics and Astronomy, Texas A\&M University, College Station, TX 77843-4242}
\affil{George P. and Cynthia Woods Mitchell Institute for Fundamental Physics and Astronomy, Texas A\&M University, College Station, TX 77843-4242}
\affiliation{Cosmic Dawn Center, Niels Bohr Institute, University of Copenhagen, R\aa dmandsgade 62, 2200 Copenhagen, Denmark\\}
\author{McDonald, M.}
\affil{MIT Kavli Institute for Astrophysics and Space Research, MIT, Cambridge, MA, O2139, USA}
\author{Muzzin, A.}
\affil{Department of Physics and Astronomy, York University, 4700 Keele St., Toronto, Ontario, M3J 1P3, Canada}
\author{Nantais, J.}
\affil{Departamento de Ciencias F{\'i}sicas, Universidad Andr{\'e}s Bello, Fern{\'a}ndez Concha 700, Las Condes, RM 7591538, Chile}
\author{Rudnick, G.}
\affil{Department of Physics \& Astronomy, University of Kansas, 1251 Wescoe Hall Drive, Malott room 1082, Lawrence, KS 66045}
\author{{\"U}bler, H.}
\affil{Cavendish Laboratory, University of Cambridge, 19 J.J. Thomson Avenue, Cambridge CB3 0HE, UK}
\affil{Kavli Institute for Cosmology, University of Cambridge, Madingley Road, Cambridge CB3 0HA, UK}
\author{van Kampen, E.}
\affil{European Southern Observatory, Karl-Schwarzschild-Str. 2, 85748, Garching bei M{\"u}nchen, Germany}
\author{Webb, T. M. A.}
\affil{Department of Physics, McGill Space Institute, McGill University, 3600 rue University, Montr{\'e}al, Qu{\'e}bec, Canada, H3A 2T8}
\author{Wilson, G.}
\affil{Department of Physics and Astronomy, University of California, Riverside, 900 University Avenue, Riverside, CA 92521, USA}
\author{Yee, H. K. C.}
\affil{The David A. Dunlap Department of Astronomy and Astrophysics, University of Toronto, 50 St George St., Toronto, ON M5S 3H4, Canada}

\begin{abstract}

We investigate the resolved kinematics of the molecular gas, as traced by ALMA in CO (2-1), of 25 cluster member galaxies across three different clusters at a redshift of $z\sim1.6$. This is the first large-scale analysis of the molecular gas kinematics of cluster galaxies at this redshift. By separately estimating the rotation curve of the approaching and receding side of each galaxy via kinematic modeling, we quantify the difference in total circular velocity to characterize the overall kinematic asymmetry of each galaxy. 3/14 of the galaxies in our sample that we are able to model have similar degrees of asymmetry as that observed in galaxies in the field at similar redshift. However, this leaved 11/14 galaxies in our sample with significantly higher asymmetry, and some of these galaxies have degrees of asymmetry of up to $\sim$50 times higher than field galaxies observed at similar redshift. Some of these extreme cases also have one-sided tail-like morphology seen in the molecular gas, supporting a scenario of tidal and/or ram pressure interaction. Such stark differences in the kinematic asymmetry in clusters versus the field suggest the evolutionary influence of dense environments, established as being a major driver of galaxy evolution at low-redshift, is also active in the high-redshift universe.

\end{abstract}

\section{Introduction}

It has been shown that the peak of the cosmic star formation rate density happened around $\sim$ 10 Gyr ago, at $z \sim 2$, \citep{Shapley+11, Madau+14}. Along with especially high star formation rates, massive ($M_\mathrm{stellar} \gtrsim 10^{10}~{\rm M}_{\odot}$) galaxies at this epoch differ from present day galaxies in other significant ways, e.g. the stellar mass-halo mass relation \citep{Behroozi+15}, the mass-size relation \citep{Perret+14}, and the mass-metallicity relation \citep{Huang+19}. Massive, star-forming galaxies at this redshift also have significantly elevated gas to stellar mass ratios \citep{Tacconi+10, Tacconi+13, Tacconi+18, Scoville+17}. Furthermore, some high-redshift galaxies in dense environments, i.e. clusters, have been found to have gas mass to stellar mass fractions even higher than field galaxies at the same redshift \citep{Noble+17, Hayashi+2018, Noble+19, Tadaki+19}. However, other studies have also found populations of high-redshift cluster galaxies with normal gas fractions as predicted from galaxy scaling relations \citep[e.g.][]{Rudnick+17, Williams+22}, or even depleted molecular gas reservoirs \citep{Alberts+22}. These diverse results highlight the need for further investigation to better understand the influence of the cluster environment at high redshift.

It has been postulated that high gas fractions in galaxies at this redshift may be the result of environmental effects, such as ram pressure from the cluster environment. A number of galaxies affected by ram pressure observed in the molecular gas phase have shown enhanced molecular gas fractions in either the disk, ram pressure stripped tail, or both, often preferentially on the leading side of the galaxy, i.e. the side experiencing maximum ram pressure \citep{Jachym+14, Jachym+17, Verdugo+15, Lee+17, Moretti+18, Cramer+20, Cramer+21}. As well as finding elevated gas fractions in a sample of $z=1.6$ cluster galaxies, \citet{Noble+19} also found several galaxies with head-tail morphology, and disturbed velocity fields. Along with environmental effects like ram pressure, cluster galaxy morphology and kinematics could also be affected by the elevated tidal interaction rates in clusters when compared to the field \citep{Fried+88, Byrd+90, Henriksen+96}. Furthermore, a significant fraction of cluster members at redshifts ranging from $z=0$ to $z=1.5$ have been accreted from galaxy groups \citep{McGee+09}, meaning that the effects of high-density environments can affect current cluster members even earlier than their accretion into the cluster - known as pre-processing \citep{Haines+15, Jaffe+15, Jaffe+16}.

One method of studying the effects of the environment on the evolution of galaxies is by studying the kinematic pattern of a galaxy, and quantifying the degree of asymmetry present. This type of analysis is most commonly done in investigating merging versus rotating galaxies, as mergers significantly disrupt the ordered rotation of a galaxy, and increase the velocity dispersion, to such a degree that it can be observed even with relatively low spectral and spatial resolution. There are a number of studies of the multiphase gas and stellar kinematics of merging galaxies at low redshift, both single-object \citep[e.g.][]{Combes+88}, and on larger scales such as the SAMI galaxy survey in H$\alpha$ \citep{Bloom+18}. There are also a number of high-redshift surveys of gas in galaxies, such as the SINS survey in H$\alpha$ \citep{Shapiro+08, FS+19}, the KMOS$^{\mathrm{3D}}$ survey in H$\alpha$ \citep{Wisnioski+15, Wisnioski+19}, and the ALPINE-ALMA survey in C[II] \citep{Jones+21}. The KMOS$^{\mathrm{3D}}$ survey resulted in a number of works exploring the kinematic properties of galaxies from $0.7 < z < 2.7$, including analyzing the rotation curve of galaxies \citep{Lang+17, Genzel+20, Price+21}, and studying the Tully-Fisher relation \citep{Ubler+17}. Specific to the cluster environment, the KMOS cluster survey \citep{Beifiori+17, Prichard+17} studied the stellar and ionized gas kinematics of cluster galaxies from $z=1.4-1.8$ to investigate the history of the evolution of clusters, but did not investigate kinematic asymmetry.

Analysis of kinematic asymmetry in low-redshift galaxies, where high resolution studies are more easily done, have shown that ram pressure with a significant edge-on wind angle component can lead to a one-sided asymmetry in the ionized gas rotation curve \citep{Boselli+94, Boesch+13}, and in some cases in the molecular gas rotation curve \citep{Lee+17, Cramer+20, Cramer+21}. In particular, \citet{Boesch+13} studied the kinematic asymmetry of the ionized gas in a large sample of both cluster and field galaxies at $z \sim 0.17$, and found the degree of kinematic distortion in clusters was on average 75\% higher than the field sample. The kinematic distortion in clusters was not correlated with stellar body distortion, suggesting cluster effects that affect gas but not stars, like ram pressure, are likely responsible for the elevated degree of kinematic asymmetry. A survey of 46 $z \sim 1.5$ galaxy cluster members with KMOS conducted by \citet{Boehm+20} quantified the kinematic asymmetry of the ionized gas rotation curves, and found elevated rates when compared to the low-redshift cluster galaxies from \citet{Boesch+13}. However, the data are relatively low resolution. When the data from \citet{Boesch+13} were degraded to the resolution of the observations in \citet{Boehm+20}, the signature of kinematic disturbance being due to hydrodynamical effects, as opposed to tidal effects, disappeared. This highlights the need for high-resolution observations in order to understand the environmentally driven evolutionary factors in high redshift clusters.

High spatial and spectral resolution ALMA studies of the kinematics of molecular gas at high redshift in both cluster and field galaxies are still rare. There is particular value in investigating the molecular gas phase, as the dense molecular gas is the site of star formation. Studying the evolution of the molecular gas phase in the cluster environment can help us to understand the process of galaxy evolution and quenching \citep{Kenney+04, Vollmer+12, Boselli+14, Lee+17, Cramer+20, Cramer+21}. At present, the only larger scale study of well-resolved (and not lensed), high-redshift ($z\sim 1.5$) cluster galaxies observed in CO is \citet{Noble+19}. A survey by \citet{Lee+19} also studied the molecular gas kinematics of 11 galaxies in a protocluster at $z \sim 2.5$, but the galaxies were barely resolved, in most cases having only a single beam across each galaxy, so an analysis of kinematic asymmetry and environmental effects would be difficult with this sample. There are high-resolution CO observations of high-redshift galaxies in the field, but they are mostly composed of single object studies \citep[e.g.][]{Genzel+13, Ubler+18, Molina+19}; small samples, such as a limited subset from the larger \citep{Genzel+20} ionized gas sample; or detected gas in lensed galaxies \citep[e.g.][]{Shen+21, Rizzo+21}, which carries additional uncertainty from the source-plane reconstruction when studying morphological and kinematic properties.

\subsection{Kinematic asymmetry in high-redshift field galaxies}

Of all observational high-redshift galaxy studies, the largest sample of quantitatively assessed velocity asymmetry comes from \citet{Genzel+20}, investigating mainly ionized gas rotation curves. The authors investigated a sample of 41 rotation curves from a combined sample of observations of galaxies from $z=0.65-2.45$, including data from SINFONI, KMOS-VLT, and CO observations from IRAM-NOEMA. Based on an analysis of the rotation curve reflection symmetry about the dynamical center for each galaxy in the sample, \citet{Genzel+20} find no convincing evidence for perturbation or environmental interaction affecting the kinematics for the large majority of the sample.  A follow-up by \citet{Ubler+21} simulated and mock observed similar galaxies at $z=2$ in the TNG50 simulation to those presented in the \citet{Genzel+20} sample, and also measured kinematic asymmetry. \citet{Ubler+21} found significantly higher disturbances to a regular velocity field, with a mean degree of asymmetry $\sim$30 times higher than that measured by \citet{Genzel+20}. \citet{Ubler+21} suggest that a reason for this huge difference in degree and rate of kinematic asymmetry in the sample is that half of the simulated galaxies they select have either high accretion rates, or are near massive companions, and could be experiencing tidal effects from close gravitational interaction. These effects are indicated by correlated, large vertical and
radial gas motions with respect to the disc plane. Ram pressure stripping from the host galaxy halo may also have an effect, as has been observed in galaxy pairs \citep{Moon+19}. In contrast, only 5/41 galaxies in the \citet{Genzel+20} sample have potentially close companions ($\Delta r=6-21$ kpc), and these companions are all low-mass, so strong interaction is more unlikely.

Rates and degrees of tidal interaction, as well as the strength of ram pressure, are significantly higher near the center of dense environments like galaxy clusters. Thus, a study of the kinematic asymmetry in clusters, where the signal is more likely to be strong, could help to answer whether effects like ram pressure and gravitational interaction are the cause of the asymmetry difference in the \citet{Ubler+21} \& \citet{Genzel+20} samples, and establish their role in gas kinematic evolution and observed elevated gas fractions in dense environments.

As such, we present an analysis of the molecular gas kinematics from ALMA observations of 3 clusters, identified in the Spitzer Adaptation of the Red-Sequence Cluster Survey (SpARCS) fields, at $z=1.6$ \citep{Muzzin+09, Wilson+09, Demarco+10}. The study we present here is the first of its kind for this gas phase in high-redshift clusters. First we present the observational data used in this work, including that from HST and ALMA, in Section 2. In Section 3, we describe our one-sided rotation curve modeling approach for each galaxy comparing the approaching and receding side. We then quantify the difference in the rotation curve of the two sides, and compare asymmetry rates and magnitudes within these clusters to surveys of high redshift galaxies in the field. Finally, in Section 4, we discuss the implications of the rates and degrees of asymmetry we find for the larger picture of galaxy evolution as a function of environment at high redshift.

Throughout this work, we assume a galactic $\alpha_{\mathrm{CO}}$ of 4.36 for converting CO to H$_2$ mass \citep{Bolatto+13}, which includes a 36\% correction for Helium. We also assume a $\Lambda$CDM cosmology with $\Omega_{\mathrm{M}}=0.3$, $\Omega_{\Lambda}=0.7$, and $\mathrm{H}_{0}=70 \mathrm{~km} \mathrm{~s}^{-1} \mathrm{Mpc}^{-1}$.

\section{Observations}

\subsection{Optical/Infrared Photometry and Spectroscopy of $z \sim 1.6$ SpARCS clusters}

The three clusters of galaxies studied in this paper, J022426–032330 (J0224), J033057–284300 (J0330) and J022546–035517 (J0225), were discovered within the 42 deg$^2$ SpARCS fields (see additional information in Table 1 of \citealt{Nantais+16}). All three clusters are spectroscopically confirmed with redshifts of $z=1.633$, $z=1.626$, and $z=1.59$ respectively \citep{Lidman+12, Muzzin+13, Nantais+16}, and contain, in total, 113 spectroscopically confirmed members. In total, there are observations over 16 bands spanning optical to NIR ($ugrizYK_{s}$ and F160W), as well as IR and FIR (3.6/4.5/5.8/8.0/24/250/350/500$\mu$m) that are used to estimate photometric redshifts for the entire cluster field \citep{Nantais+16}. The environmental quenching efficiency, based on the observed quenched fraction, is estimated to be $16 \pm 16\%$, which, although there is significant uncertainty, suggests the environmental quenching efficiency is close to 0 in these clusters at this redshift \citep{Nantais+17}.

The central cluster regions have deep HST imaging from the “See Change” program (GO-13677 and GO-14327; \citealt{Hayden+21}) in F160W with the WFC3-IR camera. The HST images shown in this paper are from this filter, and can be found in MAST: \dataset[10.17909/yzg9-mj62]{http://dx.doi.org/10.17909/yzg9-mj62}. The data were reduced using the Drizzlepac software.

\subsection{ALMA Observations}

The ALMA data used for this project consists of four separate single pointings aimed at the $z=1.6$ clusters J0224, J0225, and J0330, combining data from Cycle 5 (2017.1.01228.S, PI: Noble) and Cycle 6 (2018.1.00974.S, PI: Noble). These clusters were observed in Band 3 at 88 GHz to detect the CO(2-1) line with a spatial resolution of $\sim{0.35}\arcsec$. Integrated properties such as gas masses, gas fractions, and depletion times of the galaxies identified in these three clusters are tabulated in \citet{Noble+17}, and spatially-resolved gas sizes, kinematics, and updated gas masses for galaxies within J0225 were presented in \citet{Noble+19}. For the three pointings over clusters J0224 and J0330, in addition to the standard `full pass' data received from ALMA, we were also able to make use of a subset of `semi-pass' data. These data were taken during a period of phase discontinuities in several baselines, due to an  error with the correlator software. We utilized the baseline phase data to flag data from when the correlator failed. We then concatenated the unflagged semi-pass data with the full pass data and continued with data reduction. 

The data were combined using the CASA software package version 6.1 \citep{McMullin+07} with the \textit{tclean} routine using natural weighting to maximize the amount of emission recovered, and a channel width of 50 km s$^{-1}$. Cleaning was done with the auto-multithresh routine \citep{Kepley+20}, after manual checking of the automatically drawn clean regions, down to a low-noise threshold of $1.5\sigma$. The resulting cubes have an average rms of $\sim$0.1 mJy beam$^{-1}$ in 50 km s$^{-1}$ channels, and beam minor and major axes of $0.35-0.5$\arcsec, varying slightly between clusters.

Moment maps shown in this paper were generated using the \textit{Search} routine in 3DBarolo \citep{DiTeodoro+15}. The routine identifies high threshold peaks (in our case we specified a minimum of 4$\sigma$) and then searches for any contiguous emission down to a 2$\sigma$ level in both spatial and spectral directions in the cube. We detect 25 galaxies in total, 8 galaxies in the J0225 cluster identified in \citet{Noble+19}, as well as an additional 3 galaxies in the J0330 cluster, and 14 galaxies over two pointings in the J0224 cluster. \citet{Noble+17}, which was based on shallower, unresolved data from Cycle 3, previously identified 5 of these galaxies, one in J0330, and four in J0224. The first unresolved ALMA observations from Cycle 3 were conducted as a semi-blind CO survey of each of these three spectroscopically confirmed SpARCS clusters; pointings were chosen to optimize the number of spectroscopically-confirmed cluster members within each FOV, while also ensuring that some were undergoing dusty star formation based on 24$\mu$m detections.  Therefore, the original survey was agnostic to star forming efficiency, as there were no prior CO detections. Follow-up spatially resolved ALMA pointings from Cycle 5 \& 6 had slightly shifted FOVs to maximize the known CO detections from the Cycle 3 data. Thus, the galaxies presented here represent the strongest CO-detected cluster members within the FOV of the observations, but cover a range of stellar masses and SFRs (though typically on or around the star-forming main sequence).

\begin{table}
\centering
\begin{tabular}{cccc}
\hline
\head{1.5cm}{\textbf{Cluster ID}} & \head{2.0cm}{\textbf{Redshift}} & \head{2.0cm}{\textbf{CO detected galaxies}} & \head{1.5cm}{\textbf{Rotation modeled galaxies}} \\ \hline
J0224 & 1.63 & 14 & 9 \\
J0225 & 1.59 & 8  & 5 \\
J0330 & 1.63 & 3  & 0
\end{tabular}
\caption{\textbf{(1)} SpARCS cluster name. \textbf{(2)} Spectroscopically determined cluster redshift. \textbf{(3)} Number of ALMA CO(2-1) detected galaxies in each cluster. \textbf{(4)} Of the ALMA detected galaxies, the number of galaxies in each cluster which can be modeled with 3DBarolo. Those which cannot are either dispersion dominated, or too small, or low S/N, for modeling.}
\label{tab:cluster_props}
\end{table}

\section{Analysis}

\subsection{Kinematic modeling}

We utilized the kinematic modeling software tool 3DBarolo \citep{DiTeodoro+15} to ultimately estimate the rotation curve for each galaxy in our sample. 3DBarolo uses a 3-dimensional modeling approach to simulate a datacube that minimizes the residual function defined by the input data pixels and a chosen weighting function. The 3D modeling approach allows for rigorous treatment of beam smearing, which is especially important for these types of galaxies which have sizes of only a few resolution elements, and has been shown to provide accurate estimates of the rotation curve for these types of galaxies \citep{DiTeodoro+15}. A number of parameters governing the type of model drawn, in our case the centroid, systemic velocity, position angle, inclination, circular velocity, and velocity dispersion, can be either fixed, or left to vary as free parameters, optionally guided by an initial guess. 

Of the 25 galaxies we detect with ALMA, 16 are strong enough detections that we are able to fit a 3DBarolo model with at least 3 points of a rotation curve (see Table \ref{tab:cluster_props} for a breakdown for each cluster; note that cluster J0330 contains no modeling candidates). All would be considered rotation dominated (where the maximum rotation velocity $v_{\mathrm{max}}$ is greater than the average velocity dispersion $\sigma$) based on the 3DBarolo outputs for these parameters, although we make no strong claim on the degree to which they are rotation dominated due to the significant uncertainty on the magnitude of $v_{\mathrm{max}}$ from the uncertainty in the inclination. Of the 9 remaining galaxies we could not fit with 3DBarolo, 5 were not fit due to being too small and/or low S/N to generate a reliable model with more than a single model tilted ring due to a lack of pixels. The other 4 galaxies are more strongly detected, with sufficient pixels to, in theory, fit with 3DBarolo, but lack sufficiently strong rotation for the program to produce a fit, and thus are likely ``dispersion dominated". Two out of the 16 galaxies we were able to fit with 3DBarolo did not have a typical gas distribution. Instead of a peak in the surface brightness at the center, with falling surface brightness towards the outskirts, we only detected several isolated clumps distributed throughout the disk. While 3DBarolo produced a model of these galaxies, we found the models to be unreliable with very large error bars, and so have excluded these galaxies from the final sample of 14 galaxies with which we proceeded to the next step.

We ran each galaxy through 3DBarolo in a two-step process. In the first run, we left the ring centroid ($x_0$, $y_0$), the systemic velocity, and the position angle as free parameters. We provided an initial guess for the central ring position of the gas distribution centroid, and an initial guess of the position angle and the systemic velocity to those which best symmetrized the kinematic pattern. Because the 14 galaxies are rotation dominated, it is possible to provide a relatively accurate estimate of both the position angle (PA) and systemic velocity v$_{sys}$. We then inspected the model residual patterns for this first run to make sure they exhibited no clear patterns of mismatched centroid, PA, or v$_{\mathrm{sys}}$ (see \citet{Warner+73, Kruit+78} for examples of these visually identifiable residual patterns). The inclination $i$ of each galaxy was estimated based on a 2D gaussian fit to the molecular gas distribution, using the \textit{imfit} routine in CASA. It is very difficult to accurately estimate the inclination of high-redshift galaxies with observations at any wavelength due to the lack of resolution elements across the disk. However, for the purposes of this paper where we are comparing symmetry about the minor axis, and are not concerned with an accurate measurement of the maximum circular rotation of each galaxy, uncertainty in the inclination has no significant impact on our analysis. While the inclination does affect the shape of the tilted rings modeled with 3DBarolo, the uncertainty this would introduce is insignificant when compared to that already present from the beam smearing of the data.

After this first run to constrain the ring geometry, we then did a second run of 3DBarolo with these geometric parameters fixed, and the rotation velocity $V_{\mathrm{rot}}$ and velocity dispersion $\sigma_{\mathrm{gas}}$ were allowed to vary as free parameters. We set a lower limit for the velocity dispersion of $25$ km s$^{-1}$ to match what we estimate to be the minimum velocity dispersion we could measure in a moment 2 map based on the channel width of our data cubes of 50 km s$^{-1}$. 

This two-step fitting process with 3DBarolo helps to ensure a more realistic rotation curve measurement than if the geometric and kinematic parameters were fit at the same time, which tends to lead to more unphysical discontinuities in the rotation field due to oscillations in the inclination and the position angle  \citep{DiTeodoro+15, Su+22}. In the second run we also utilize 3DBarolo's built-in method for estimating the uncertainty in the measurement of the rotation velocity and velocity dispersion. This is done after the residual fit function is minimized via a Monte-Carlo method. 3DBarolo calculates a range of models centered around this minimum by a series of random Gaussian draws, that allow for oversampling this parameter space. The residuals in this region usually behave as a quadratic function, and the errors are estimated as the range where this quadratic function shows a residual increase of 5\%. See \citet{DiTeodoro+15} for more details.

\begin{figure*}
	\plotone{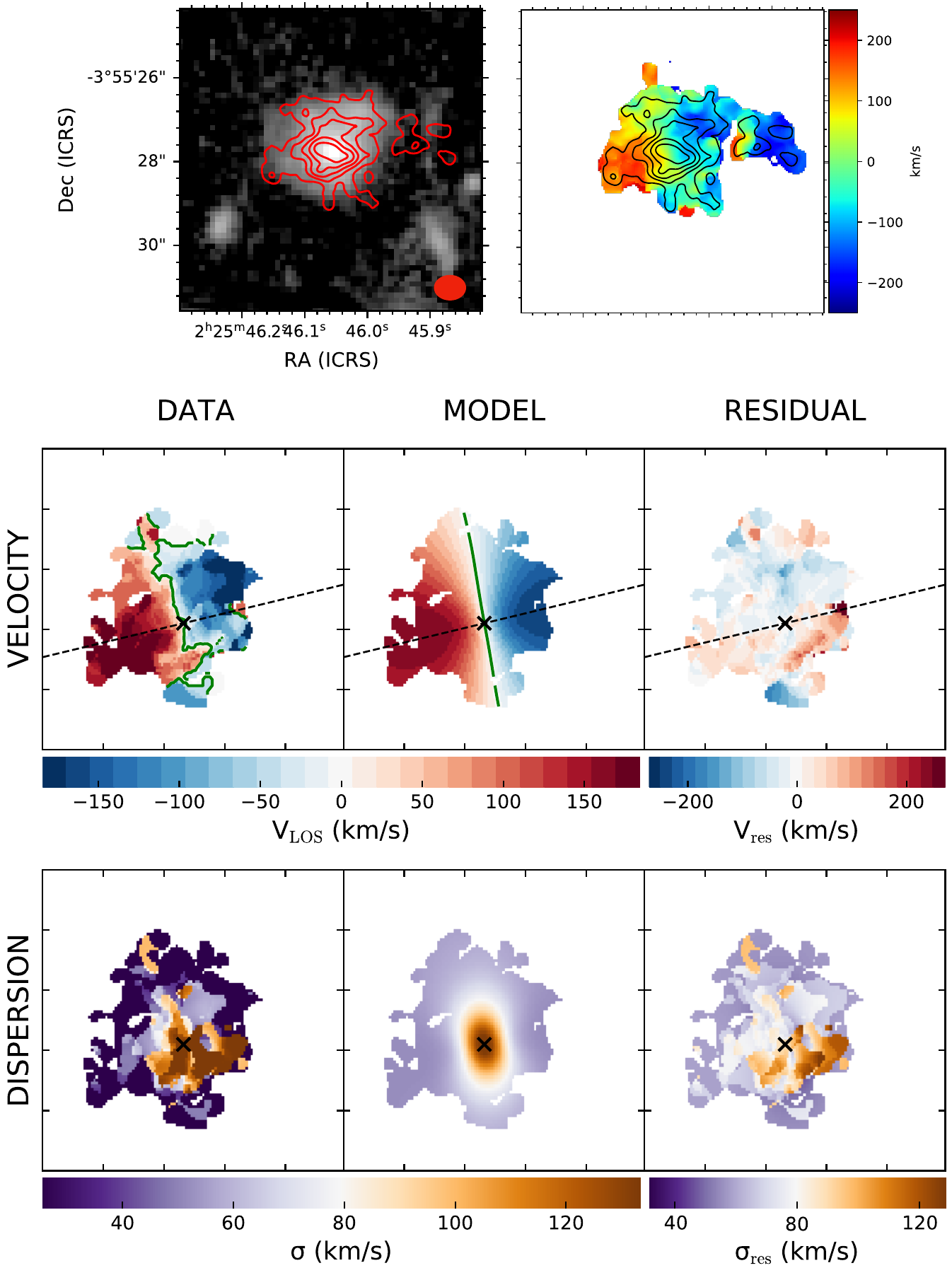}
	\caption{\textbf{Top:} On the left is an HST image of the galaxy J0225-371 in the F160W filter. Overlayed in red contours is the ALMA CO(2-1) moment 0 (intensity) map, with contour levels of 2$\sigma$, 3$\sigma$, 4$\sigma$, 5$\sigma$, 6$\sigma$. The beam size is shown with the red ellipse. On the right is a moment 1 (velocity) map, with the same moment  contours. \textbf{Middle:} A moment 1 map, showing the data on the left, the model generated with 3DBarolo in the middle, and the residual from subtracting the data from the model on the right. The cross indicates the model-determined kinematic center coordinates, the green line shows the central velocity, and the dotted line shows the major axis. \textbf{Bottom:} A moment 2 (velocity dispersion) map, model, and residual image.}
	\label{fig:moment_example}
\end{figure*}

\begin{figure*}
	\plotone{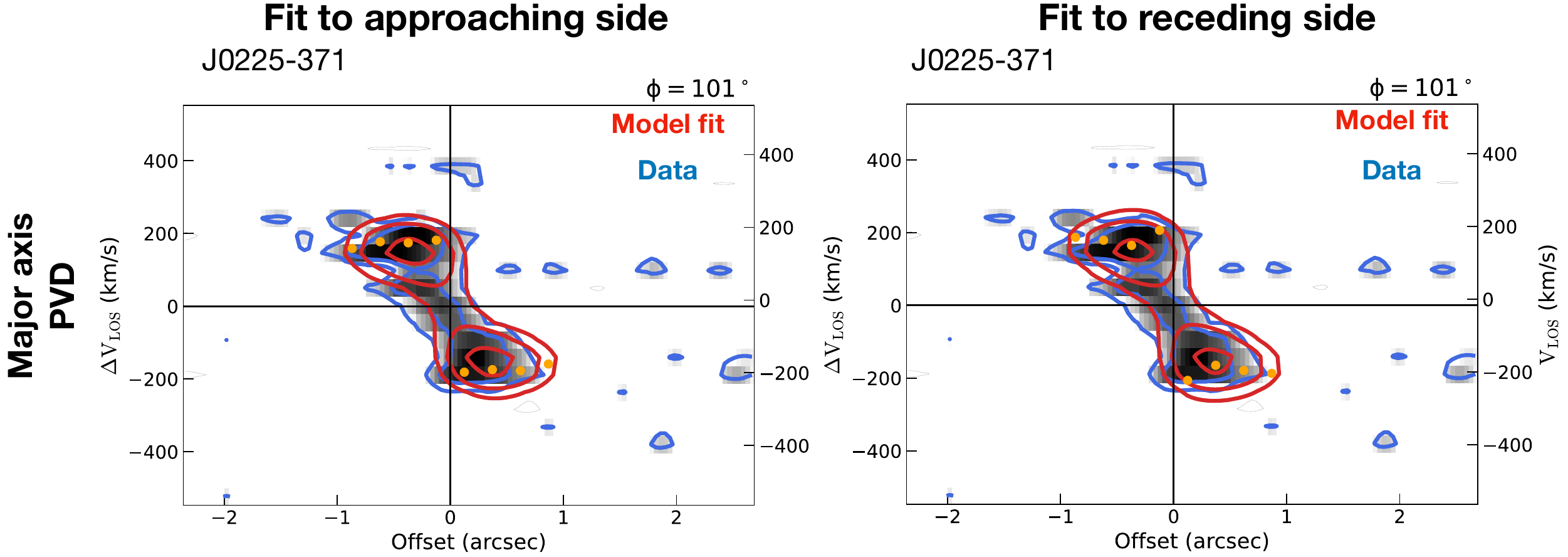}
	\caption{A comparison of the circular velocity fit to the approaching side on the left, and the receding side on the right, of J0225-371. The plots show a PVD encompassing the full span of the galaxy through the major axis. The yellow points show the circular velocity predicted by the modeling with 3DBarolo at each ellipse, spaced by 0.25 arcec. The blue contours show a smoothed version of the data, shown in gray. The red contours show a model fit only to one side of the galaxy. Both contours are shown with levels of 2$\sigma$, 3$\sigma$, 4$\sigma$, 5$\sigma$. This galaxy shows highly axisymmetric velocity structure.}
	\label{fig:PVD_371}
\end{figure*}

\begin{figure*}
	\plotone{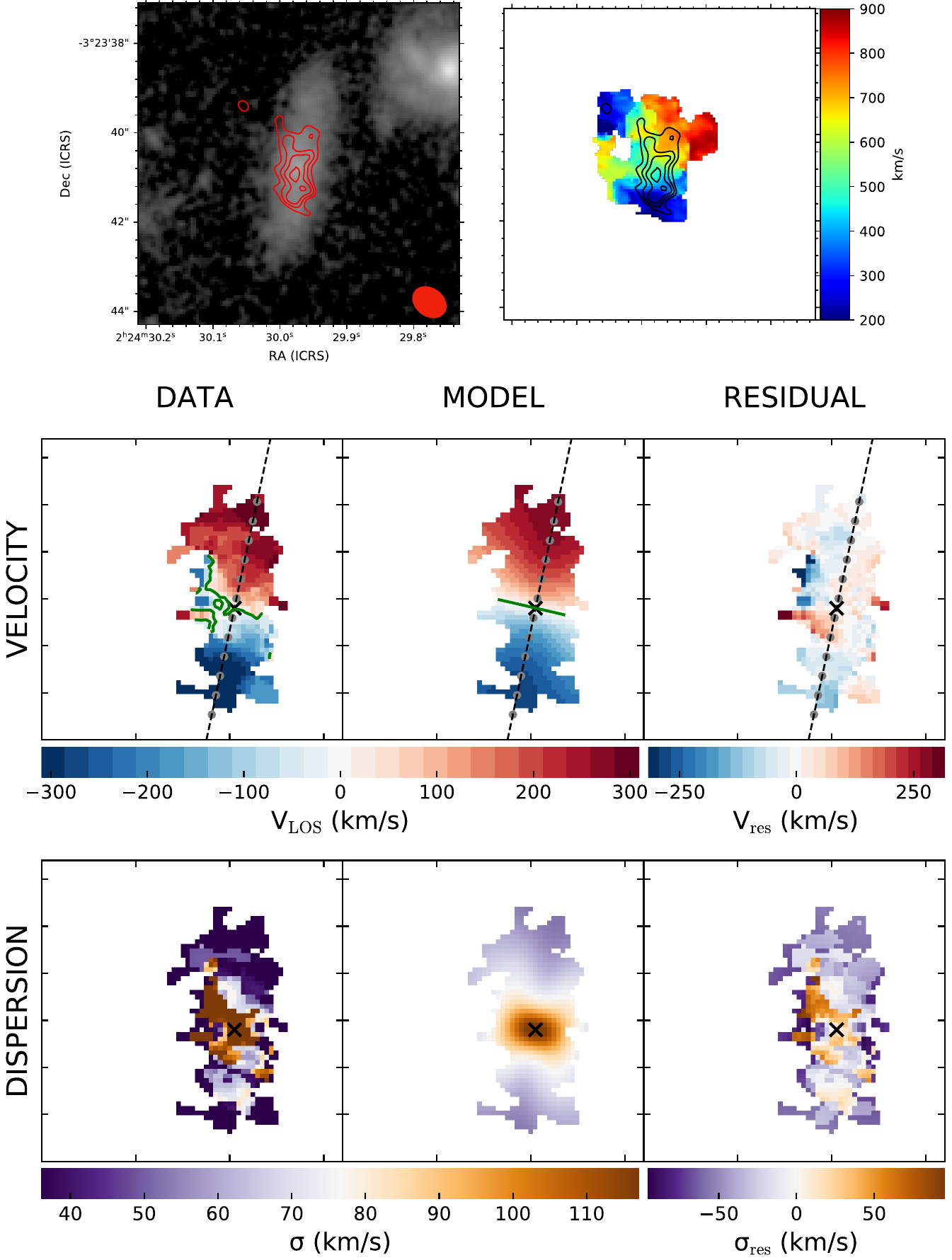}
	\caption{\textbf{Top:} On the left is an HST image of the galaxy J0224-244 in the F160W filter. Overlayed in red contours is the ALMA CO(2-1) moment 0 (intensity) map, with contour levels of 2$\sigma$, 3$\sigma$, 4$\sigma$, 5$\sigma$, 6$\sigma$. The beam size is shown with the red ellipse. On the right is a moment 1 (velocity) map, with the same moment  contours. \textbf{Middle:} A moment 1 map, showing the data on the left, the model generated with 3DBarolo in the middle, and the residual from subtracting the data from the model on the right. The cross indicates the center coordinates, the green line shows the central velocity, and the dotted line shows the major axis. \textbf{Bottom:} A moment 2 (velocity dispersion) map, model, and residual image.}
	\label{fig:moments_244}
\end{figure*}

\begin{figure*}
	\plotone{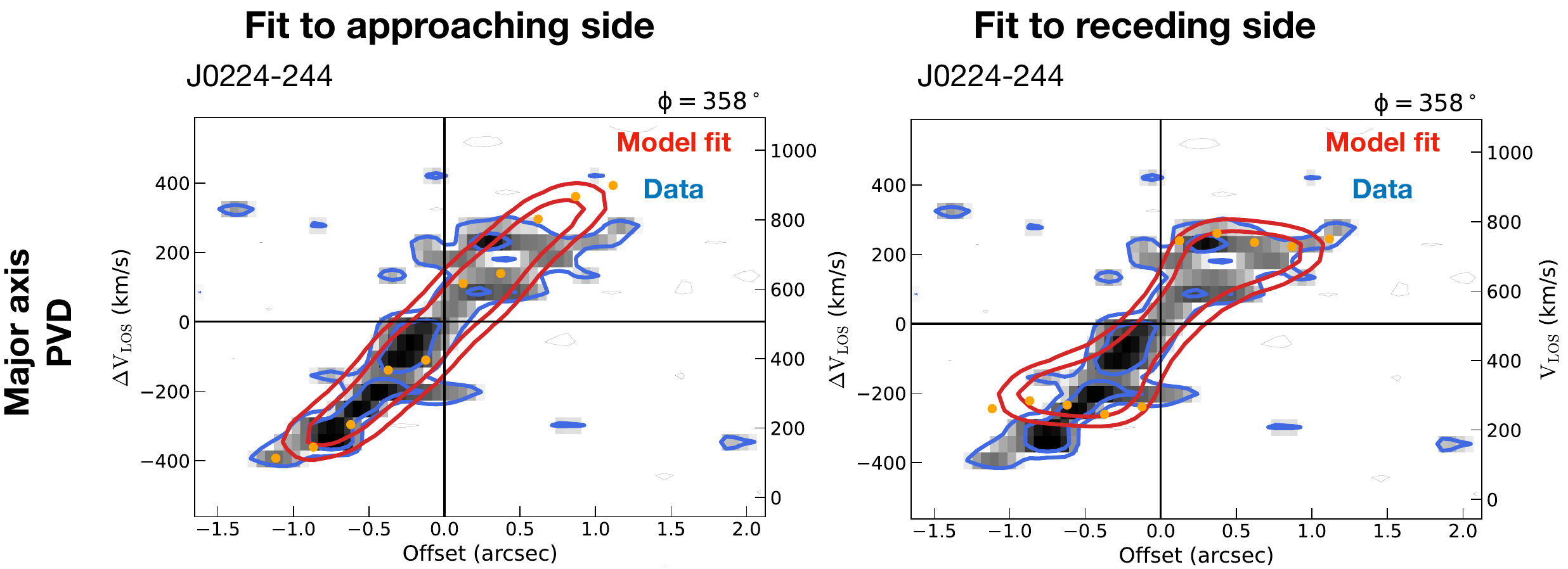}
	\caption{A comparison of the circular velocity fit to the approaching side on the left, and the receding side on the right, of J0224-244. The plots show a PVD encompassing the full span of the galaxy through the major axis. The yellow points show the circular velocity predicted by the modeling with 3DBarolo at each ellipse, spaced by 0.25 arcec. The blue contours show a smoothed version of the data, shown in gray. The red contours show a model fit only to one side of the galaxy. Both contours are shown with levels of 2$\sigma$, 3$\sigma$, 4$\sigma$, 5$\sigma$. In contrast with J0225-371, this galaxy shows non-axisymmetric velocity structure.}
	\label{fig:moment_combo_weight}
\end{figure*}

\subsection{Measuring kinematic asymmetry}

In Figure \ref{fig:moment_example} we show an example of the kinematic modeling results for a moment 1 velocity map, and moment 2 velocity dispersion map, displaying our data, the model, and the residual image. The galaxy shown is J0225-371, studied previously as part of the sample analyzed in \citet{Noble+19}. The galaxy is relatively well fit by an axisymmetric model, as evidenced by the relatively low residuals, with no clear residual pattern indicative of an incorrect choice for fitting parameters. There is a feature to the SW side of the galaxy that has high velocity dispersion, and a consistent positive residual that may be due to a non-circular motion in the disk from a structure like a spiral arm. A position-velocity diagram (PVD) encompassing all the CO emission oriented along the major axis of this galaxy (see Figure \ref{fig:PVD_371}) shows the galaxy is relatively symmetric. A model produced by a fit to the approaching side and a model produced by a fit to the receding side have very little discernible difference. In contrast, a galaxy like J0224-244, shown in Figure \ref{fig:moments_244}, has an easily visible large difference for a model fit to only the approaching versus only the receding side (Figure \ref{fig:moment_combo_weight}). While the difference in degree of asymmetry between J0225-371 and J0224-244 are clearly visible when inspecting the PVD, we seek a method to quantify the degree of asymmetry of these galaxies. We consider two different methods for this purpose.

\subsubsection{Asymmetry quantization}

The first method is to simply calculate the sum over ring radius of the offset of points on the rotation curve on the approaching side ($V_{\mathrm{A}}$) from the receding side ($V_{\mathrm{R}}$) normalized by the errors $\sigma$ and the number of points $n$, as shown in Equation \ref{velocity_eq}. This equation is similar to the equation for calculating the reduced chi-squared parameter for goodness of fit. We do this for both our sample, and the sample of simulated galaxies from \citet{Ubler+21}, provided to us by the authors. The errors ($\sigma$) for our sample are from the uncertainty in the rotation velocity produced by 3DBarolo via bootstrapping. For the \citet{Ubler+21} sample they are estimated via mock observation of sources with realistic injected noise:

\begin{equation}
    A_{\text {vel}}=\sum_{r} \frac{(V_{\text {norm, A}_r} - V_{\text {norm, R}_r})^2}{n\sqrt{\sigma_{\text {norm, A}_r}^2+\sigma_{\text {norm, R}_r}^2}}
\label{velocity_eq}
\end{equation}

\begin{figure*}
	\plotone{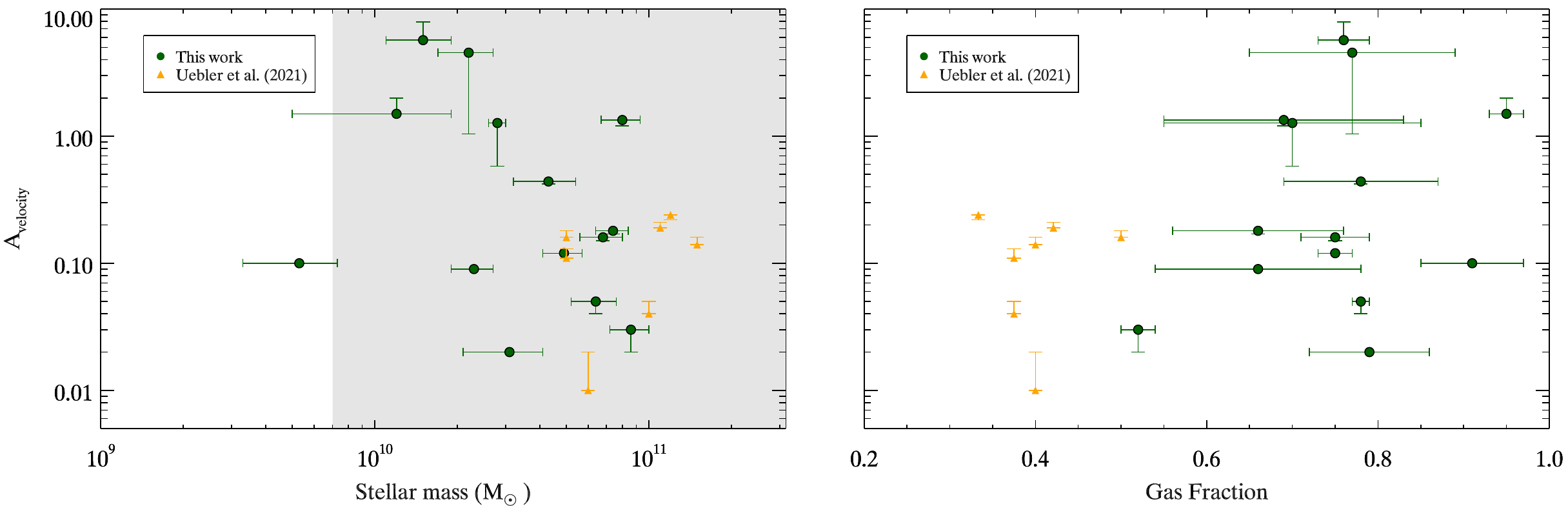}
	\caption{\textbf{Left:} A comparison of the stellar mass and the total velocity difference $A_{\text {vel}}$ (defined in Equation 1) measured between the sampled points on the velocity curve of the approaching and receding sides of each galaxy in our sample in the two clusters J0224 and J0225 (green), and the simulated galaxies from \citet{Ubler+21} (orange). The stellar mass range of the sample from \citet{Genzel+20} is shown in the grey shaded region. \textbf{Right:} A comparison of the gas fraction (M$_{\mathrm{gas}}$ / (M$_{\mathrm{gas}}$ + M$_{\mathrm{*}}$) and the total velocity difference $A_{\text {vel}}$ for both our sample and the simulated sample from \citet{Ubler+21}. As noted in \citet{Noble+19}, the gas fractions of the cluster galaxies in our sample are especially high when compared to those predicted from standard scaling relations. All velocity curves have been normalized to a maximum of 1 to remove any influence from the inclination estimate. The error in $A_{\text {vel}}$ is estimated by removing the outermost (which is  also the lowest S/N) ring fit by 3DBarolo, and recalculating the asymmetry.}
	\label{fig:dist_diff}
\end{figure*}

If a galaxy had a perfectly symmetric rotation curve about the minor axis, the difference between points on the approaching and receding side would be 0. The larger the offset, and smaller the relative error, the larger the $A_{\text {vel}}$ parameter will be. For this method, we normalize each rotation curve such that the maximum velocity is set to 1, so that the highly uncertain inclination estimated from 3DBarolo is not a factor in the calculation. We also calculate $A_{\text {vel}}$ for the set of 7 simulated galaxies from the \citet{Ubler+21} sample. We display the results in Figure \ref{fig:dist_diff}. The stellar masses for the galaxies in our sample were estimated from SED modeling using CIGALE \citep{Boquien+19}, using the 16 bands spanning optical to FIR available for SpARCS galaxies. We used the implementation of the \citet{Dale+14} dust model within CIGALE.

To test whether $A_{\text {vel}}$ is correlated with stellar mass, and whether it is correlated with gas fraction, we ran a Spearman's rank correlation test on each data sample. Considering only our data on $A_{\text {vel}}$ and stellar mass, we find a Spearman correlation value of $\rho = -0.34$ (considered a `weak' correlation), and a p-value of 0.23. If we include the \citet{Ubler+21} data as well, we find a stronger correlation, with $\rho = -0.43$ (considered a `moderate' correlation), and a p-value of 0.05. In contrast, considering only our data on $A_{\text {vel}}$ and gas fraction, we find a Spearman correlation value of $\rho = 0.10$ (considered a `very weak' correlation), and a p-value of 0.73. The inclusion of the \citet{Ubler+21} data somewhat increases the degree of correlation, resulting in a Spearman correlation value of $\rho = 0.23$ (considered a `weak' correlation), and a p-value of 0.30. Therefore, while there is statistical evidence for a monotonic relation between lower stellar mass and increased asymmetry, there is little evidence for a correlation with gas fraction. Further sampling of this parameter space in future surveys could greatly increase our certainty in these results.

\begin{figure*}
	\plotone{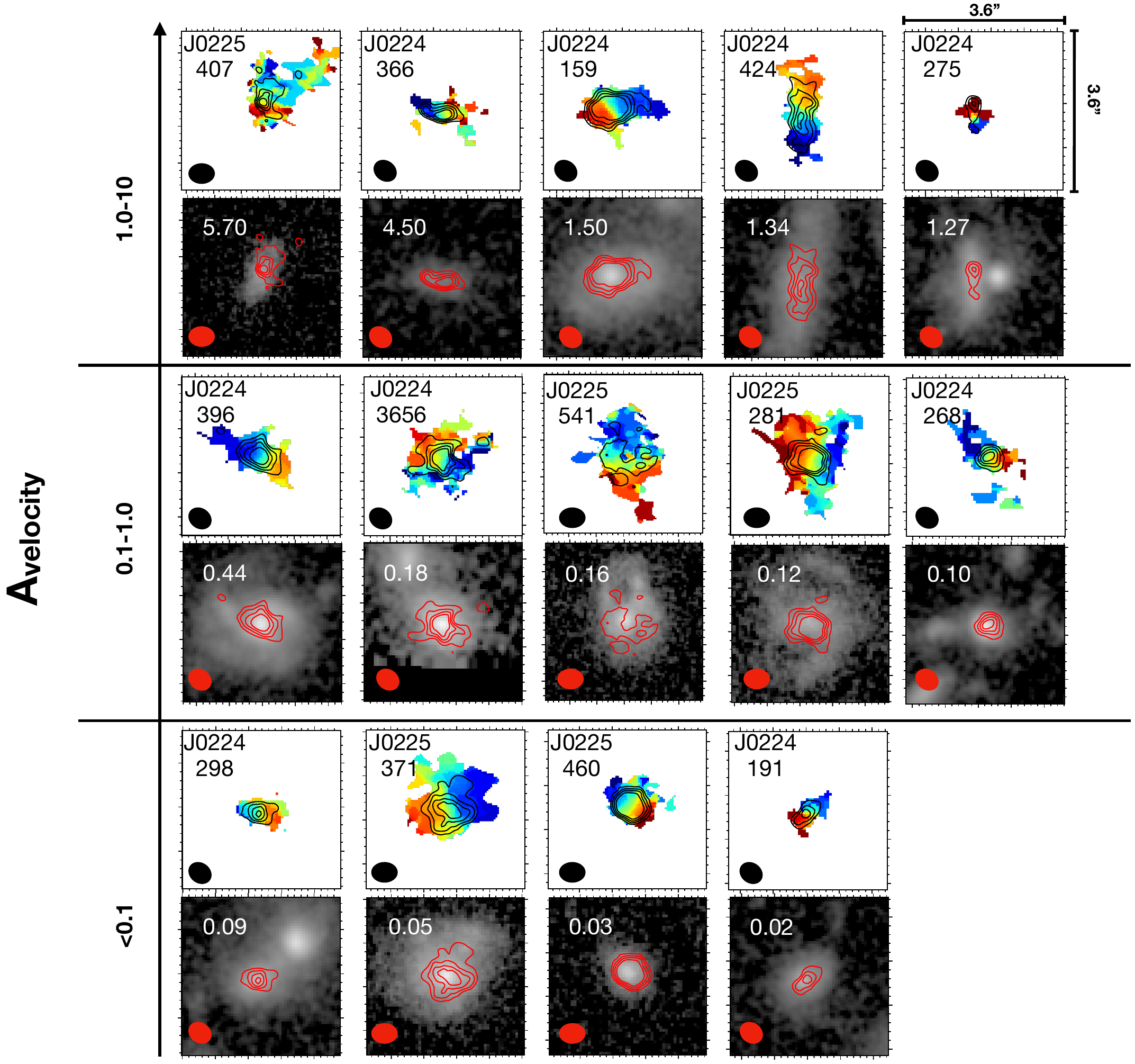}
	\caption{In this plot, we show moment 1 velocity maps with moment 0 contours in black (contour levels of 2$\sigma$, 3$\sigma$, 4$\sigma$, 5$\sigma$), and HST images with moment 0 contours in red, for each galaxy. Each stamp has dimensions of 3.6\arcsec $\times$ 3.6\arcsec ($\sim 30$ $\times$ $30$ kpc). The cluster each galaxy is found in as well as its catalogue ID is shown in black text on the moment 1 map.  We order the galaxies from highest $A_{\text {vel}}$ at the top left, to lowest $A_{\text {vel}}$ at the bottom right. The $A_{\text {vel}}$ value is printed in white text on the moment 0 \& HST overlay cutout. Note that the average $A_{\text {vel}}$ of the \citet{Ubler+21} sample we compare to in Figure \ref{fig:dist_diff} is $\sim$0.1.}
	\label{fig:mom_asym}
\end{figure*}


We find that overall our sample has a mean $A_{\text {vel}}$ of $\sim$1.1, and a median of $\sim$0.2, while the \citet{Ubler+21} sample has a mean $A_{\text {vel}}$ of $\sim$0.1, and a median of 0.1. On a case by case level, about half our sample has a similar range of $A_{\text {vel}}$ as the \citet{Ubler+21} galaxies. However, as indicated by the very large difference in the mean, several galaxies (preferentially those of low mass) in our sample are much more extremely asymmetric than any in the \citet{Ubler+21} sample. We discuss the implications of this further in Section 4. In Figure \ref{fig:mom_asym} we show the 14 galaxies in our sample ordered from highest to lowest $A_{\text {vel}}$.

\subsubsection{$\Delta \chi^2_{\mathrm{red}}$ method for asymmetry quantization}

\begin{figure*}
	\plotone{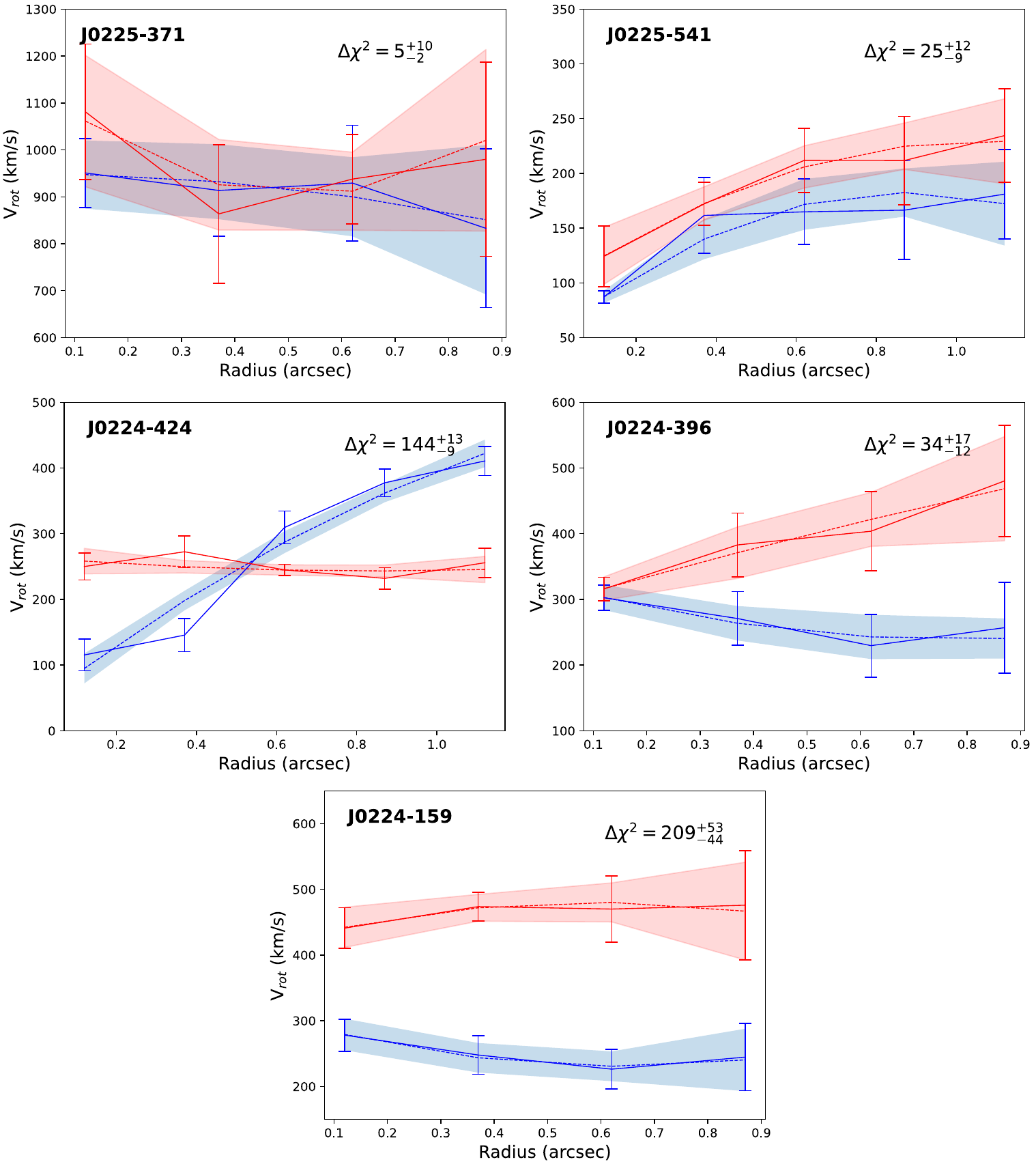}
	\caption{A sample of the rotation curves of the approaching (blue) and receding (red) side of each galaxy in our sample, showing only galaxies with more than three points on the rotation curve on each side. The solid line shows the rotation velocity estimate at each ring by 3DBarolo. The dotted line is a parabola fit to the data. The shaded regions show the 1$\sigma$ uncertainty of the parabolic fit. Printed in the figure legend is the total $\Delta \chi^2_{\mathrm{red}}$ value for comparing each fit to its opposite side. The lower and upper limits on the value of $\Delta \chi^2_{\mathrm{red}}$ are based on recalculating this parameter from the 1$\sigma$ uncertainty of the parabolic fit.}
	\label{fig:velocity_curve}
\end{figure*}

\begin{figure*}
	\plotone{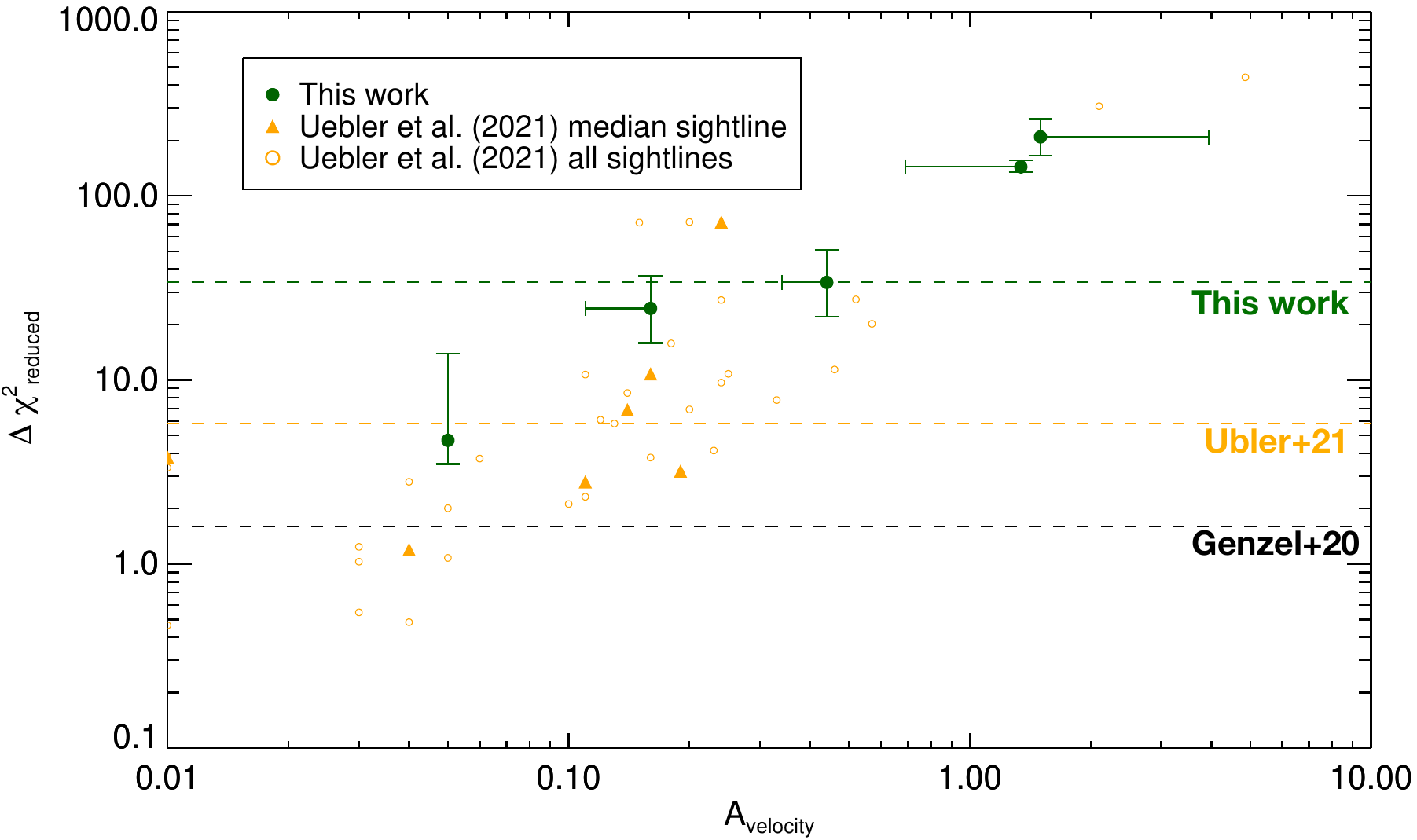}
	\caption{Here we show in green the calculated values of the A$_{\mathrm{vel}}$ and $\Delta \chi^2_{\mathrm{red}}$ parameters for measuring velocity asymmetry described in Section 3.3 for each of the five galaxies in our sample for which it could be measured. Error bars for $\Delta \chi^2_{\mathrm{red}}$ are estimated based on the uncertainty from the curve fitting to the data. The error in $A_{\text {vel}}$ is estimated by removing the outermost (which is  also the lowest S/N) ring fit by 3DBarolo, and recalculating the asymmetry. The green dashed line shows the median $\Delta \chi^2_{\mathrm{red}}$ for our sample. The orange points show a comparison between the A$_{\mathrm{vel}}$ and $\Delta \chi^2_{\mathrm{red}}$ from \citet{Ubler+21}. The orange open circles show the the 5 sightlines mock observed for each of the 7 galaxies, while the filled triangles show the median value of the 5 sightlines for each galaxy. Overall, we see a linear trend between A$_{\mathrm{vel}}$ and $\Delta \chi^2_{\mathrm{red}}$. The orange and black dotted lines show the median of $\Delta \chi^2_{\mathrm{red}}$ for the galaxies in the \citet{Ubler+21} and \citet{Genzel+20} samples respectively.}
	\label{fig:chisq_reg}
\end{figure*}

We also consider a second method for comparison to the results from this technique used by both \citet{Genzel+20} \& \citet{Ubler+21}. This method involves fitting a parabolic function to the approaching side, and another parabolic function to the receding side, and then comparing that same fit to the opposite side of the galaxy and calculating the difference in goodness-of-fit, the reduced chi-squared value, between the two sides as $\Delta \chi^2_{\mathrm{red}}$. This method is useful for smoothing over any discontinuities in the rotation curve which could be due to the uncertainty in the measurement of the rotation velocity, assuming that a smooth rotation curve is a better representation of the true rotation curve. However, this method is of limited use for our sample as only 5/16 galaxies have more than 3 points on the rotation curve on both the receding and approaching side. At least four points are needed for the number of degrees of freedom to be large enough to calculate a reduced chi squared. We present the results of this chi-squared fitting method for our sample in Figure \ref{fig:velocity_curve}. Overall we find that the $\chi^2_{\mathrm{red}}$ fit to each single side across the sample has a mean value of $\sim$1.1, indicating the parabolic fit is a good match to the general velocity trend as a $\chi^2_{\mathrm{red}}$ value of about one generally indicates. We also calculate a p-value range of $0.90-0.95$ for these five galaxies, based on the number of rings fit (i.e. the number of degrees of freedom) and the chi squared value. We plot the $\Delta \chi^2_{\mathrm{red}}$ value of each of the five galaxies in our sample, as well as the average $\Delta \chi^2_{\mathrm{red}}$ from the \citet{Ubler+21, Genzel+20} galaxies in Figure \ref{fig:chisq_reg}.

We calculate an average, and median, $\Delta \chi^2_{\mathrm{red}} = 83^{+21}_{-15}$ and 34, for our sample of 5 galaxies. In contrast, \citet{Ubler+21} found an average and median value of $\Delta \chi^2_{\mathrm{red}}=46.8$ and 5.8, and they also calculated an average value and median of $\Delta \chi^2_{\mathrm{red}}=3.7$ and 1.6, for a subset of 12 massive galaxies ($M_*>4 \times 10^{10} M_{\odot}$) at $z > 1.5$ in the \citet{Genzel+20} sample. This indicates that galaxies in our sample are more similar, in terms of the degree of kinematic asymmetry, to the simulated galaxies from \citet{Ubler+21}, and much more asymmetric than the galaxies in the \citet{Genzel+20} sample.

We find there appears to be a linear correlation between $A_{\text {vel}}$ and $\Delta \chi^2_{\mathrm{red}}$ for galaxies in our sample where both could be calculated (Figure \ref{fig:chisq_reg}) and, therefore, extrapolating the $A_{\text {vel}}$ comparison to the $\Delta \chi^2_{\mathrm{red}}$ results is supported by this correlation. Given that $\Delta \chi^2_{\mathrm{red}}$ for the \citet{Ubler+21} data was so much greater than that for the \citet{Genzel+20} sample, and since the $\Delta \chi^2_{\mathrm{red}}$ we calculate for our restricted sample is even greater than the \citet{Ubler+21} sample, we would expect the mean and median $A_{\text {vel}}$ for the \citet{Genzel+20} sample to be significantly less than that calculated for our sample and the \citet{Ubler+21} sample.

\begin{figure*}
	\plotone{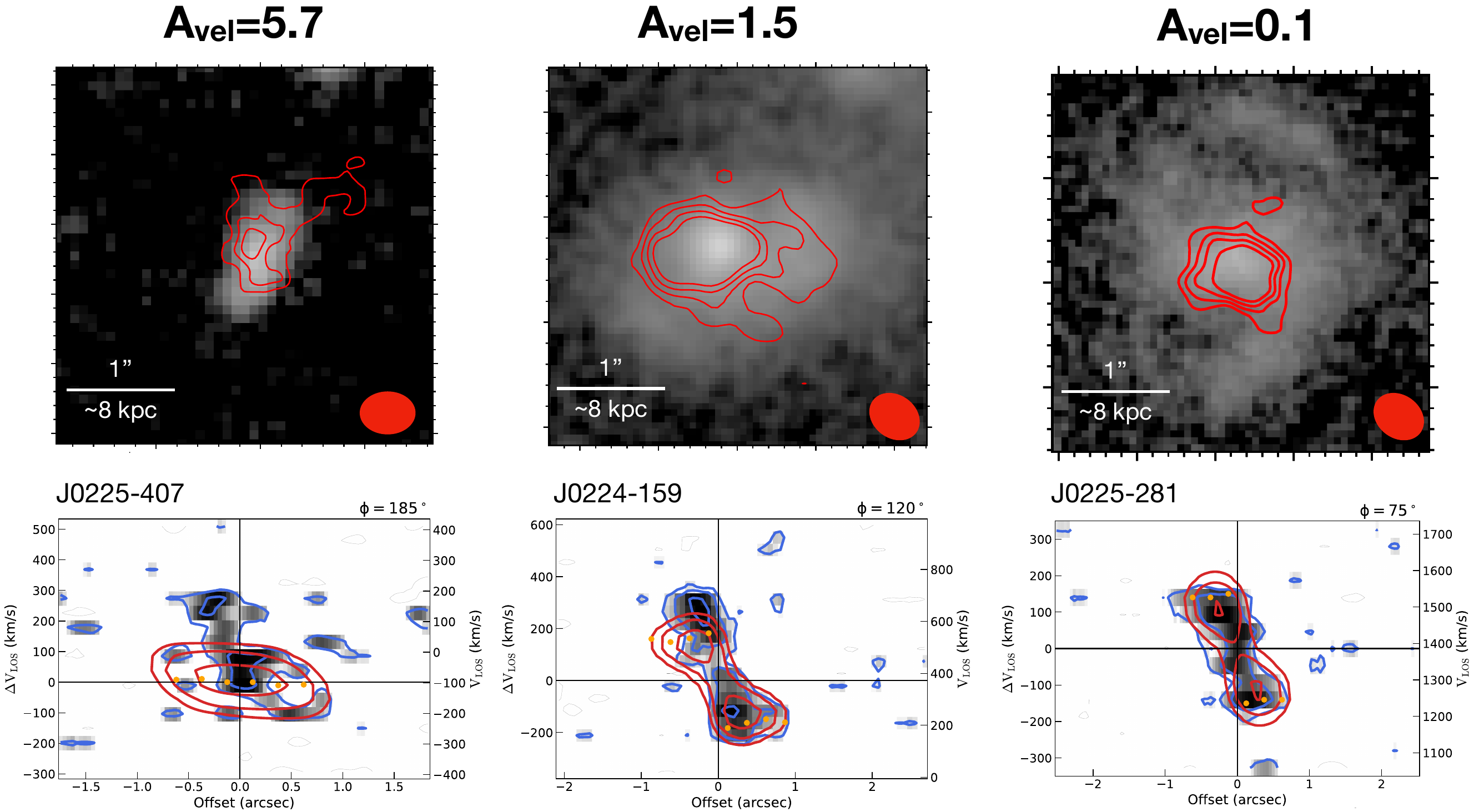}
	\caption{\textbf{Top:} An HST image of the galaxies J0225-407, J0224-244, and J0224-159, each in the F160W filter. Overlayed in red contours is the ALMA CO(2-1) moment 0 (intensity) map, with contour levels of 2$\sigma$, 3$\sigma$, 4$\sigma$, 5$\sigma$, 6$\sigma$. \textbf{Bottom:} Examples of the model produced by a circular velocity fit to only one side of each galaxy. The plots show a PVD encompassing the full span of the galaxy through the major axis. The yellow points show the circular velocity predicted by modeling with 3DBarolo at each ellipse, spaced by 0.25\arcsec. The blue contours show a smoothed version of the data (shown in black) for comparison with the red contours, which show the model. Both contours are shown with levels of 2$\sigma$, 3$\sigma$, 4$\sigma$, 5$\sigma$, 6$\sigma$. In this figure we show an example of two galaxies (left and middle) with strong asymmetries, and head-tail morphology seen in the moment 0. Galaxy J0225-407 was also presented in \citet{Noble+19} as a possible ram pressure stripped candidate based on its molecular gas morphology. We show a moment map of this galaxy generated with \textit{mommaps} which better shows the tail than the 3DBarolo generated map. On the right we show a galaxy with a more symmetric PVD and a low measured asymmetry, for comparison. Filled ellipses shown in the bottom right of each plot in the top row indicate the size and orientation of the clean beam.}
	\label{fig:tail_galaxies}
\end{figure*}

\subsection{Caveats}

At present, there are a very limited number of studies of galaxy gas kinematic asymmetry. The full parameter space of potential conditions that could affect asymmetry, including redshift and gas phase, is quite unexplored. As such, the closest samples we have with which to compare our sample are observations and simulations of mostly ionized gas rotation curves in field galaxies from $z \sim 1-2$ from \citet{Genzel+20} and \citet{Ubler+21}. Given that our data consist of molecular gas observations of cluster galaxies at redshift of $z=1.6$, there are many potential sources of systematic uncertainty that could affect this comparison. Here, we address the potential effects of some of these systematics. However, it is clear that we cannot truly understand all the potential sources of uncertainty such as gas phase, redshift, environment, halo mass, etc., without a diversity of further surveys of galaxy kinematics.

\subsubsection{Ionized vs. molecular gas kinematics}

The \citet{Genzel+20} sample is mostly composed of rotation curves derived from observations of ionized gas, and \citet{Ubler+21} attempts to simulate a similar gas phase. Thus, our direct comparison of asymmetries in the molecular gas to mainly ionized gas could introduce additional uncertainty. Molecular gas is expected to be dynamically colder than ionized gas, and several studies have shown these two gas phases have different velocity dispersion in low and high redshift cases \citep{Levy+18, Uebler+19, Girard+21}. However, there is evidence from the literature that supports that for the purposes of an asymmetry comparison, this may not significantly affect this comparison. \citet{Genzel+13, Ubler+18} found excellent agreement in the the rotation curve derived from H$\alpha$ and CO in two star forming galaxies at $z \sim 1.5$. Also, in a larger sample of 17 nearby galaxies from the ALMaQUEST survey, the authors compared the rotation curve derived with 3DBarolo from separate ionized and molecular gas observations of each galaxy \citep{Su+22}. They found that about half the galaxies had a measurable difference in the circular velocity of the two gas phases, with a median total velocity difference of 6.5 km s$^{-1}$. Furthermore, inspection of the CO-H$\alpha$ velocity curves for each galaxy in the ALMaQUEST subsample do not show significant asymmetric differences as a function of distance from the galaxy center.

While it is unknown how much the difference in CO-H$\alpha$ velocity curves may vary over cosmic time or with environment, a difference like that from the \citet{Su+22} sample is insignificant compared to the other sources of uncertainty we already take into account. Furthermore, certain environmental effects that are particularly strong in dense environments, like ram pressure, are expected to have an even greater impact on the kinematics of the less dense ionized phase than the denser molecular gas, as has been observed, for example, in \citet{Boselli+94}. Thus the kinematic asymmetry of some of the cluster galaxies studied here may in fact be an underestimate of the ionized gas asymmetry we could expect to measure with similar observations as \citet{Genzel+20}.

\subsubsection{Signal-to-Noise effects}

Another potential concern is whether the asymmetry we measure particularly in the outer regions is majorly biased by the signal-to-noise (S/N) of the observations, as opposed to real gas kinematics. Does the asymmetry calculated with the method described in this work rise as the surface brightness drops towards the outskirts of galaxies? Our analysis supports that this is not likely to be the case. First, while \citet{Genzel+20} does not indicate the S/N limit to which they measure rotation curves, there is no apparent increase in asymmetry towards the ends of the rotation curves they derive (see Figure 4 of \citet{Genzel+20}). Furthermore, inspection of a sample of the rotation curves shown in Figure \ref{fig:velocity_curve} reveals no overall trend towards an increase in asymmetry at the edges of the surface brightness limit of galaxies in our sample. All galaxies in our subset of 14 that we have modelled with 3DBarolo have the peak S/N at the innermost model ring, and drop in S/N towards the outer model rings. As seen in Figure \ref{fig:velocity_curve}, galaxy J0224-396 shows an increasing velocity curve from the second ring onward on only one side of the galaxy, a pattern that could be consistent with environmental effects that act from the outside-in more strongly on one side of the galaxy, such as ram pressure stripping \citep{Lee+17, Cramer+20}. Other galaxies like J0224-424 show a completely different velocity pattern from the inside to the outskirts, not supported by a scenario in which asymmetry increases as S/N decreases.

Finally, for extra assurance, we recalculate the asymmetry for each galaxy in our sample, excluding the outermost (lowest S/N) ring. The result (indicated by the error bars in Figure \ref{fig:dist_diff}) is only a slight drop in the mean $A_{\text {vel}}$, from 1.1 to 1.0. While this decrease could be due to removing the lowest S/N outer ring where the velocity uncertainty is highest, it could also be due to the outermost ring having the highest rotation curve asymmetry, as found by \citet{Cramer+20} in a molecular gas study of a ram pressure stripped Virgo cluster galaxy. Either way, it is a small enough decrease ($\sim 8\%$) that it is unlikely the high asymmetry is driven primarily by the S/N of our observations.

\subsubsection{Velocity curve extent}

Finally, we note that about half of the galaxies in the \citet{Genzel+20} sample have rotation curve measurements out to up to two times the radius than we do. They also, likely due to this, reach the flat or slightly falling part of the rotation curve in almost all the galaxies in their sample, while in contrast, it is difficult to tell with our data if we have reached this same radius in most of our galaxies. Furthermore, it is not necessarily clear whether any of the galaxies in our sample have a flat outer rotation curve, especially since environmental effects like tidal forces and ram pressure stripping can affect the shape of the rotation curve, especially at the outskirts \citep{Boselli+94, Lee+17, Cramer+19}. The functional form of the rotation curve could affect the overall shape of the parabolic function fit for measuring $\Delta \chi^2_{\mathrm{red}}$ in our sample compared to \citet{Genzel+20}.

By inspecting the rotation curves in Figure 4 of \citet{Genzel+20}, it is clear that if they had measured all galaxies to about the same radius as we measure ($\sim$1\arcsec) there would not be any significant difference in the measured mean asymmetry of their sample. Outside-in effects like harassment, tidal stripping, and ram pressure stripping could be behind the large difference in mean asymmetry behind our sample and the \citet{Ubler+21} sample. If so, when compared to \citet{Genzel+20}, we would expect, if we could measure gas kinematics out to a similar radius as they did, to see an even greater increase in the mean asymmetry of our sample. This would be assuming that the gas disks in galaxies we observe are not already truncated by environmental effects as has been observed in a number of low-redshift cluster galaxies \citep[e.g.][]{Chung+09, Cortese+11, Zabel+22}.

Finally, the resolution of our data is also lower than that of the \citet{Genzel+20} sample in some cases. However, \citet{Feng+20} found in a large scale analysis of kinematic asymmetry of galaxies from the MaNGA survey that variations in the spatial resolution had no discernible effect on the measured degree of kinematic asymmetry. While this is based on analysis of galaxies from a redshift range of $0 < z < 0.2$, it at least partially supports that the resolution difference between our data and the \citet{Genzel+20} sample may not affect the overall results. Furthermore, the \citet{Genzel+20} sample itself spans a redshift range of $0.65 < z < 2.5$ and a resolution range of 0.25\arcsec - 0.8\arcsec. While there is little to no kinematic asymmetry measured in any of the galaxies in the sample, they find no correlation between resolution and asymmetry.

\section{Discussion}

Overall, we find about half the galaxies in our sample (7/14) have a similar $A_{\text {vel}}$ value as galaxies in the simulated \citet{Ubler+21} sample. Of the five galaxies in our sample for which we could measure $\Delta \chi^2_{\mathrm{red}}$, one galaxy has a value ($\Delta \chi^2_{\mathrm{red}} = 3.7$) that falls within the range of $\Delta \chi^2_{\mathrm{red}} = 1-5$ measured in the \citet{Genzel+20} sample. However, our galaxies are on average more asymmetric than either the \citet{Ubler+21} sample of simulated galaxies, or the \citet{Genzel+20} sample of field galaxies. \citet{Genzel+20} found that only 3/41 galaxies in their sample had significant asymmetries in the rotation curve. We find that half of the galaxies in our sample have a $A_{\text {vel}}$ higher than any galaxy in the \citet{Ubler+21} sample, which has an average $\Delta \chi^2_{\mathrm{red}}$ more than ten times higher than the \citet{Genzel+20} sample. This suggests that kinematic asymmetry is much more common in cluster galaxies. Furthermore, there are a number of galaxies in our sample with extreme degrees of $A_{\text {vel}}$, more than ten times higher than any galaxy in the \citet{Ubler+21} sample.

Our results could be explained by environmental effects that become more extreme as the density of the environment increases, in particular, tidal interactions and ram pressure stripping. The \citet{Genzel+20} sample of field galaxies has the lowest measured asymmetry. The \citet{Ubler+21} sample are much more asymmetric than the \citet{Genzel+20} sample. While comparing simulation and observation directly can be difficult to interpret, \citet{Ubler+21} note that several of the simulated galaxies they study are in close proximity to massive galaxies. Thus, it is possible that tidal interactions and/or ram pressure effects from the companion galaxy's circumgalactic medium may contribute to the higher velocity asymmetry. Finally, our sample probes the highest density environment, clusters, where we find galaxies with degrees of asymmetry far beyond any galaxies in the field simulated by \citet{Ubler+21} or observed by \citet{Genzel+20}.

The gas distribution of some of these galaxies helps to support that environmental dependent effects could be responsible for the extreme degrees of kinematic asymmetry we measure. Some of these galaxies have asymmetric distributions of gas that could be consistent with tidal or ram pressure stripped tails (see Figure \ref{fig:tail_galaxies}), not seen in any of the galaxies in the \citet{Ubler+21} sample. At the current observation depth and resolution, it is difficult to establish whether the morphology of these galaxies is due to tidal or ram pressure interactions. At low redshift, where higher resolution is available, an asymmetric distribution of molecular gas, where the side opposite the tail often has elevated molecular gas fractions, is often a signature of ram pressure \citep{Lee+17, Cramer+20, Cramer+21}. A lack of disturbance of the stellar disk can also rule out tidal interaction or mergers, although with the currently available optical data on this sample it is difficult to determine whether this is the case. However, we note that the PVDs of 12/14 of these galaxies show only a single velocity component in PVD space at all offsets, and a strong rotation. This makes them generally inconsistent with being classified as ``mergers" as identified by \citet{Jones+21} in a similar survey of the morphology and ionized gas kinematic profiles of high-redshift galaxies. The other 2/14 galaxies may be mergers, but the PVD is complex to interpret (see Appendix A for individual PVDs of each galaxy). The galaxy J0224-3656 has two strong components at the same offset, separated significantly in velocity space, but has a relatively normal overall rotation curve. J0224-366 has a complex kinematic pattern and an irregular stellar body, and so is likely to be a recent merger or major accretion.

Furthermore, we note that a reason for the difference in degree of asymmetry in our sample and that of \citet{Ubler+21} could be that we probe down to a lower stellar mass range ($M_*=10^{9.5}~{\rm M}_{\odot})$ than \citet{Ubler+21}. It has been well documented that low-mass galaxies in clusters have higher HI-deficiencies as a result of ram pressure stripping \citep{Cortese+11}. The results of a Spearman rank correlation test also support a `moderate' monotonic correlation between decreasing stellar mass and increasing asymmetry. Moreover, the \citet{Genzel+20} sample of field galaxies has no trend with stellar mass and asymmetry, supporting that this relation may be the result of an environmental density dependent effect. A future rigorous study of the stellar body of galaxies in our sample could help us to identify or rule out tidal interaction, and establish the degree to which the high rate of asymmetry could be due to ram pressure.

We also note that \citet{Noble+17, Noble+19} found particularly high gas fractions in galaxies in these same $z=1.6$ clusters. High gas fractions in the disks and the tails of jellyfish galaxies have been found at low redshift as well \citep{Jachym+17, Moretti+18, Moretti+20}. If the cause of the elevated kinematic asymmetry rates in clusters we observe is primarily due to ram pressure, we might expect to see a correlation between the degree of kinematic asymmetry and the gas fraction. However, as seen in Figure \ref{fig:dist_diff}, and as illustrated by the results of a Spearman rank correlation test showing `very weak' correlation between gas fraction and kinematic asymmetry, this does not appear to be the case. A larger sample size across a larger range of clusters may help to determine with more certainty whether such a trend exists.

\section{Conclusion}

In summary, we have conducted the first large-scale characterization of the kinematic asymmetry of the molecular gas of high-redshift ($z=1.6$) cluster galaxies. Through an analysis of the kinematic asymmetry, conducted with the use of the kinematic modeling software 3DBarolo \citep{DiTeodoro+15}, we have modeled the rotation curve and quantified the degree of kinematic asymmetry in each of the 14 galaxies in our sample. When compared to field galaxies at similar redshift from observations \citep{Genzel+20} and simulation \citep{Ubler+21} we find some galaxies in our sample with similar asymmetry values. However, 12/14 of the galaxies in our sample have higher asymmetry than the average galaxy in the \citep{Genzel+20} sample, and 7/14 have higher asymmetry than the average galaxy in the \citep{Ubler+21} sample. Furthermore, several galaxies have kinematic asymmetry to a significantly higher degree than any galaxies in either of the reference samples. These galaxies have asymmetry values at least 10 times higher than any galaxy from \citet{Ubler+21}, and at least 50 times higher than any galaxy in the \citet{Genzel+20} sample. In some cases, they also have visible head-tail gas morphology suggestive of tidal and/or ram pressure interaction (as first noted in \citealt{Noble+19}), both of which are more common in the inner regions of dense environments like clusters when compared to the field. This is the first evidence of such an effect on molecular gas velocity fields of galaxies in clusters at high redshift.

While these results shed light on the importance of environmental effects on the evolution of molecular gas in galaxies in a novel regime, i.e. the high-redshift Universe, we stress that this study only scratches the surface. Large-scale kinematic asymmetry studies of galaxies in low-redshift clusters are still very limited. In the future, a comparison of the rate and degree of kinematic asymmetry, and how it changes with cluster redshift, mass, and evolutionary stage, would be very important to our understanding of the evolution of galaxies in clusters across cosmic time.



\acknowledgments

We thank Enrico Di Teodoro for help in optimizing the 3DBarolo software package for our data. We also thank Delaney Dunne for insights and discussion on kinematic modeling software.

This work was supported in part from HST program number GO-16300. A.N. additionally acknowledges support from the Beus Center for Cosmic Foundations at Arizona State University, and from the NSF through award SOSPA7-025 from the NRAO. GW gratefully acknowledges support from the National Science Foundation through grants AST-1517863 and AST-2205189, and from HST program numbers GO-15294 and GO-16300. Support for program numbers GO-15294 and GO-16300 was provided by NASA through grants from the Space Telescope Science Institute, which is operated by the Association of Universities for Research in Astronomy, Incorporated, under NASA contract NAS5-26555. JN is supported by Universidad Andres Bello internal grant DI-07-22/REG. MC acknowledges support from NSF grant AST-1815475. R.D. gratefully acknowledges support by the ANID BASAL projects ACE210002 and FB210003.

This paper makes use of the following ALMA data: ADS/JAO.ALMA\#2017.1.01228.S, 2018.1.00974.S. ALMA is a partnership of ESO (representing its member states), NSF (USA) and NINS (Japan), together with NRC (Canada), MOST and ASIAA (Taiwan), and KASI (Republic of Korea), in cooperation with the Republic of Chile. The Joint ALMA Observatory is operated by ESO, AUI/NRAO and NAOJ. The National Radio Astronomy Observatory is a facility of the National Science Foundation operated under cooperative agreement by Associated Universities, Inc. Based on observations made with the NASA/ESA Hubble Space Telescope, and obtained from the Hubble Legacy Archive, which is a collaboration between the Space Telescope Science Institute (STScI/NASA), the Space Telescope European Coordinating Facility (ST-ECF/ESA) and the Canadian Astronomy Data Centre (CADC/NRC/CSA).

\software{CASA (v6.1.0; \citet{McMullin+07}), Python maskmoment (\url{https://github.com/tonywong94/maskmoment}) IDL mommaps (\url{https://github.com/tonywong94/idl\_mommaps}), 3DBarolo \citep{DiTeodoro+15}}

\clearpage

\bibliography{Bibliography}{}
\bibliographystyle{aasjournal}

\clearpage

\appendix

\restartappendixnumbering

\section{PVD and asymmetry measurements}

Here we show the moment maps generated by 3DBarolo of each galaxy in our sample, as well as a PVD of the data and the model cube made from fitting the rotation velocity and velocity dispersion of only the approaching and only the receding side. The galaxies are ordered from highest to lowest $A_{\text {vel}}$ (as defined in Equation \ref{velocity_eq}), the same order as shown in Figure \ref{fig:mom_asym}.

\begin{figure*}
    \epsscale{0.85}
	\plotone{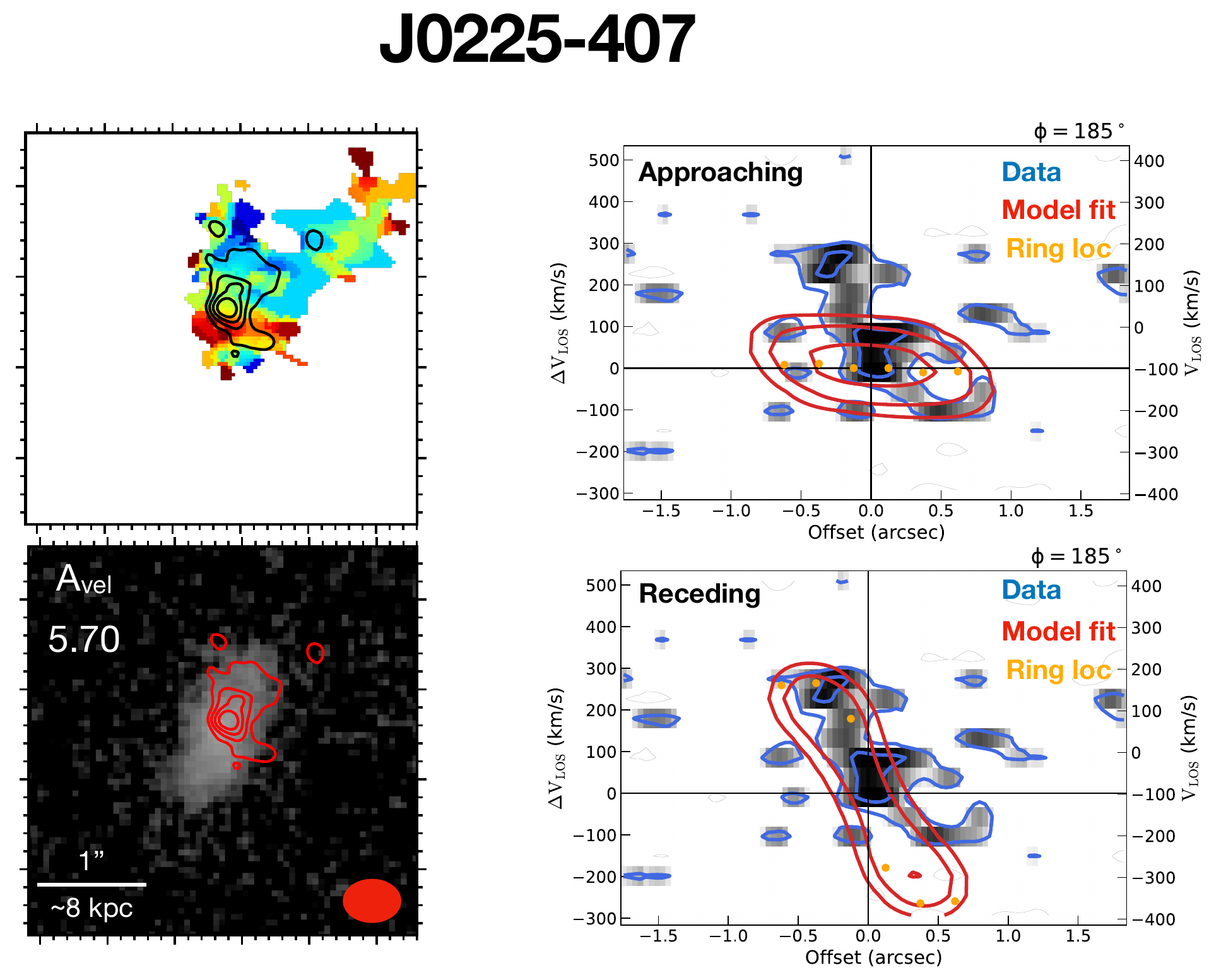}
	\caption{\textbf{Top Left:} A moment 1 (velocity) map produced by the 3DBarolo \textit{Search} routine, with the moment 0 (intensity) map, also from 3DBarolo, overlayed in black contours. Contour levels are 2$\sigma$, 3$\sigma$, 4$\sigma$, and 5$\sigma$. \textbf{Bottom Left:} An HST image in the F160W filter, with the moment 0 map from 3DBarolo overlayed in red contours. Contour levels are 2$\sigma$, 3$\sigma$, 4$\sigma$, and 5$\sigma$. The beam size is shown with the red ellipse. \textbf{Top Right:} A PVD encompassing the full span of the galaxy through the major axis. The yellow points show the circular velocity predicted by modeling with 3DBarolo at each ellipse, spaced by 0.25 arcec, fit only to the \textbf{approaching} side of the galaxy. The blue contours show a smoothed version of the data (shown in black) for comparison with the red contours, which show the model. Both contours are shown with levels of 2$\sigma$, 3$\sigma$, 4$\sigma$, 5$\sigma$. \textbf{Bottom Right:} Same as top right but fit only to the \textbf{receding} side of the galaxy.}
	\label{fig:appendix_pvd_asym}
\end{figure*}

\begin{figure*}
    \epsscale{0.85}
	\plotone{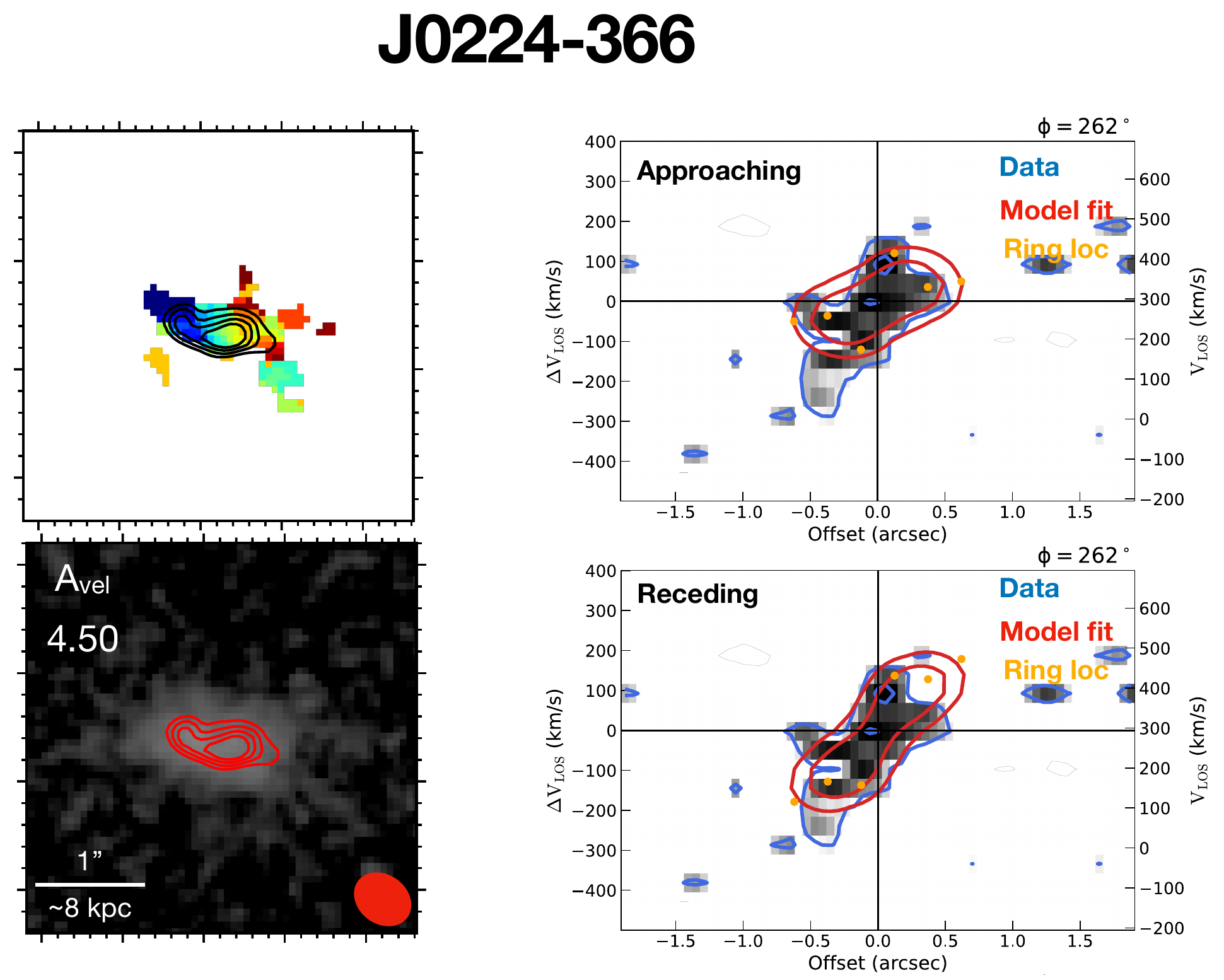}
	\caption{Caption same as Figure \ref{fig:appendix_pvd_asym}.}
	\label{fig:additional}
\end{figure*}

\begin{figure*}
    \epsscale{0.85}
	\plotone{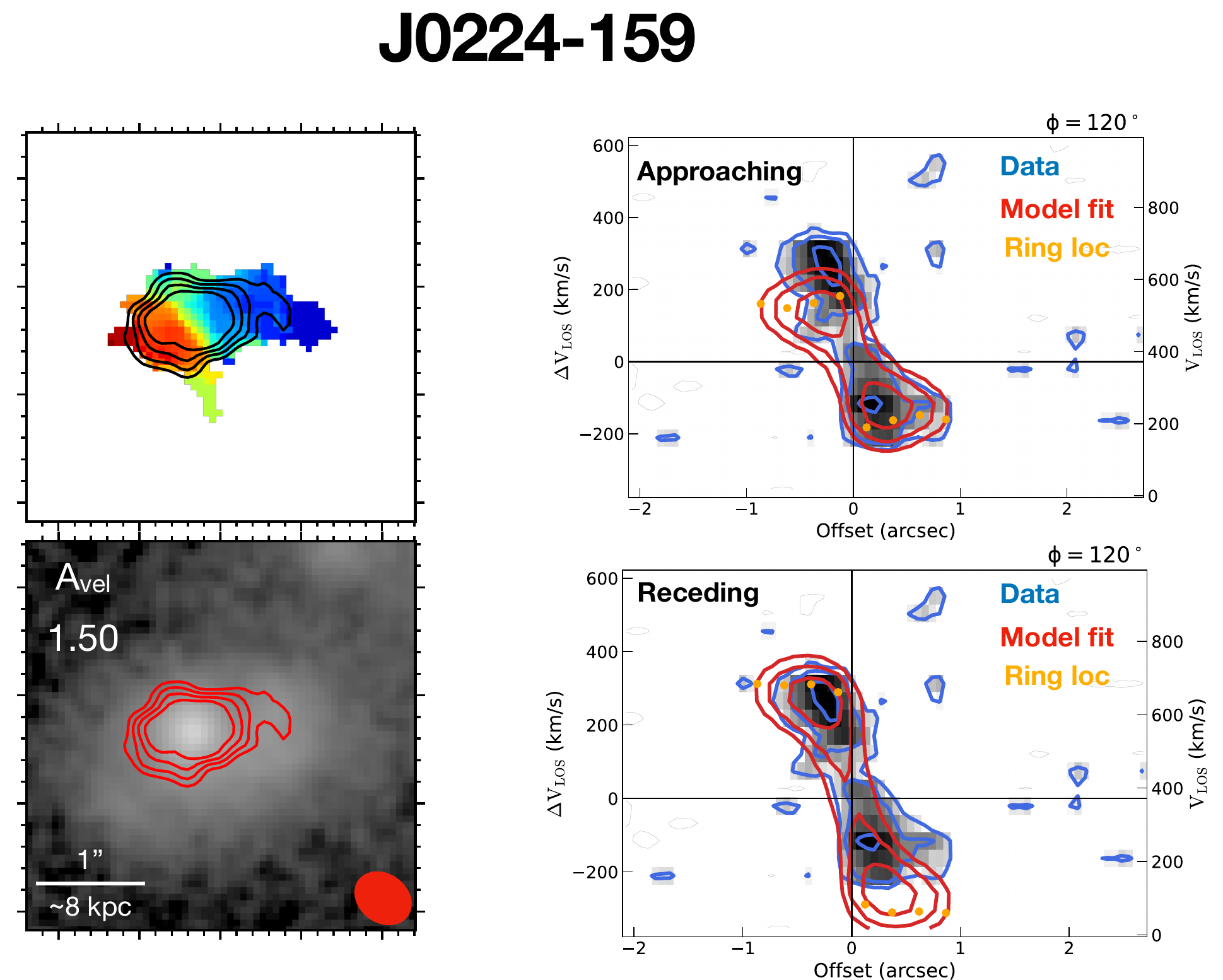}
	\caption{\textbf{Top Left:} Caption same as Figure \ref{fig:appendix_pvd_asym}.}
	\label{fig:additional}
\end{figure*}

\begin{figure*}
    \epsscale{0.85}
	\plotone{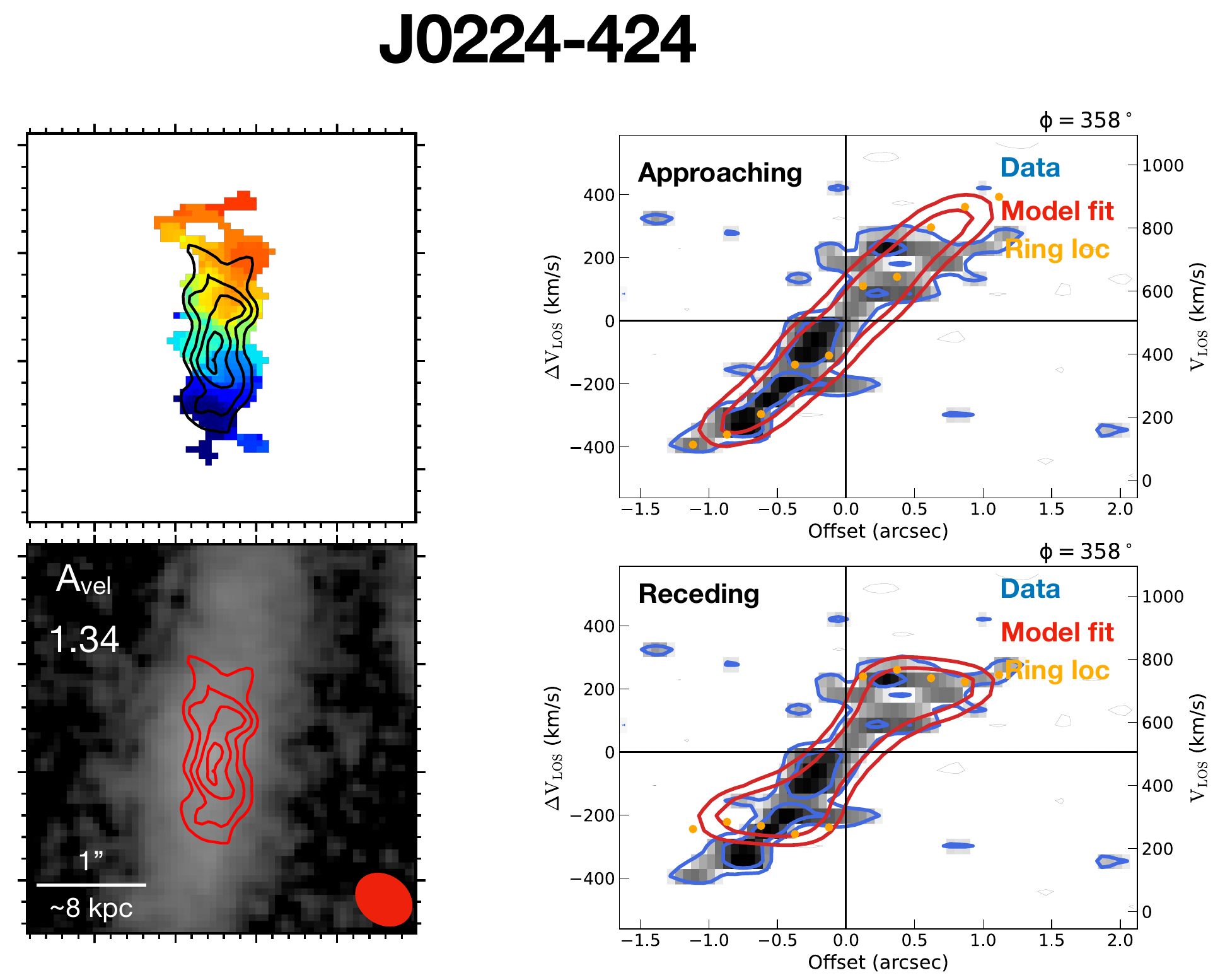}
	\caption{\textbf{Top Left:} Caption same as Figure \ref{fig:appendix_pvd_asym}.}
	\label{fig:additional}
\end{figure*}

\begin{figure*}
    \epsscale{0.85}
	\plotone{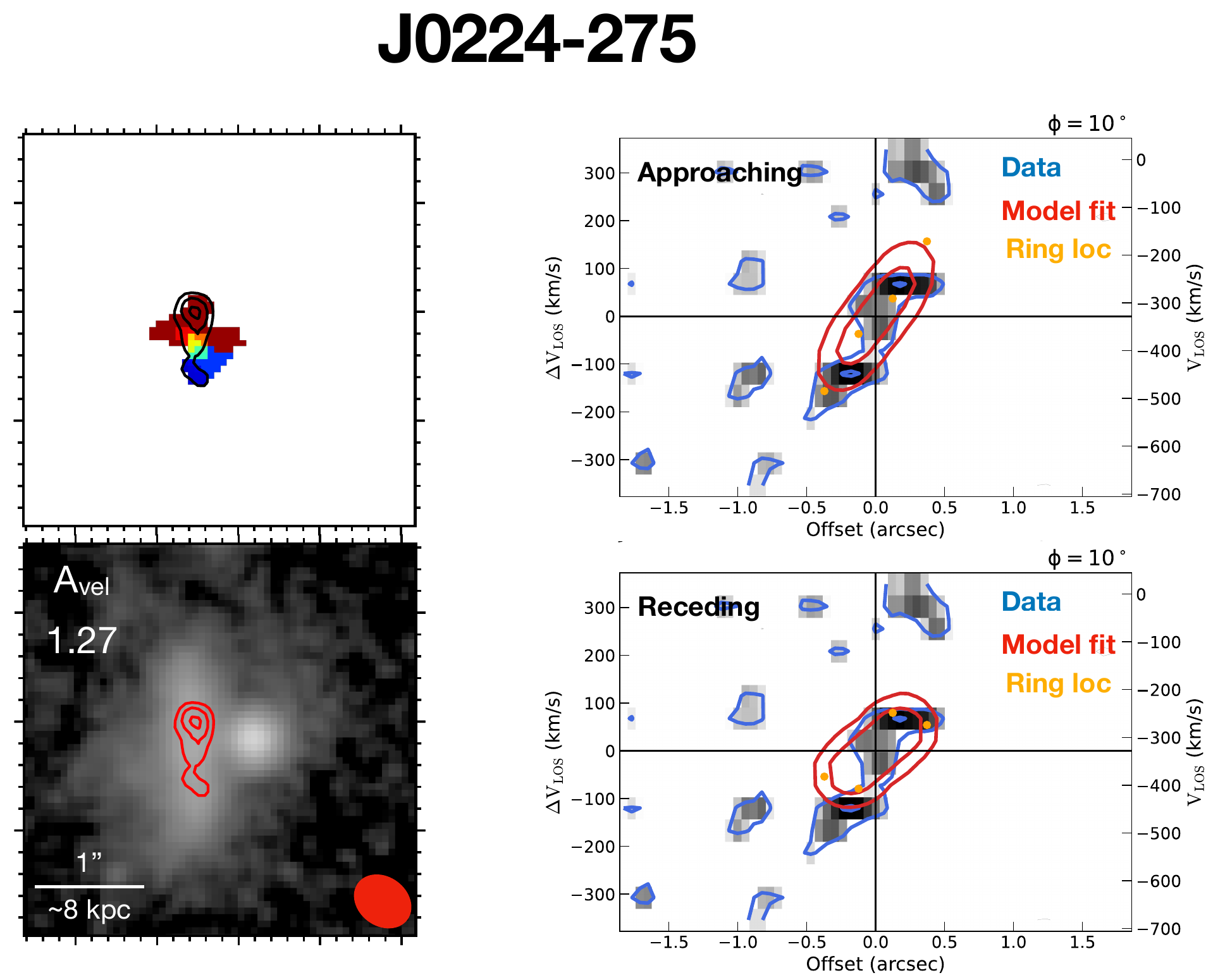}
	\caption{\textbf{Top Left:} Caption same as Figure \ref{fig:appendix_pvd_asym}.}
	\label{fig:additional}
\end{figure*}

\begin{figure*}
    \epsscale{0.85}
	\plotone{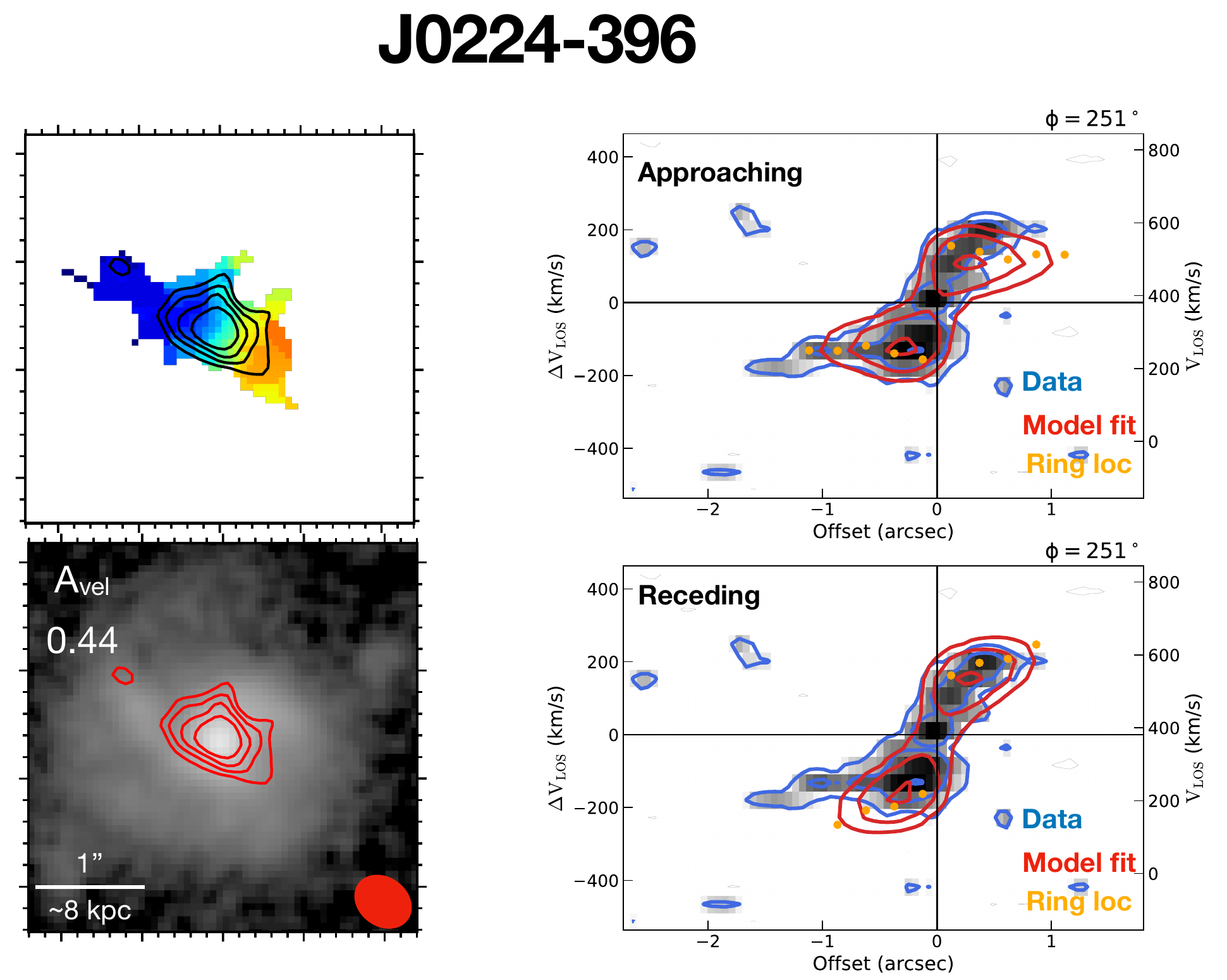}
	\caption{\textbf{Top Left:} Caption same as Figure \ref{fig:appendix_pvd_asym}.}
	\label{fig:additional}
\end{figure*}

\begin{figure*}
    \epsscale{0.85}
	\plotone{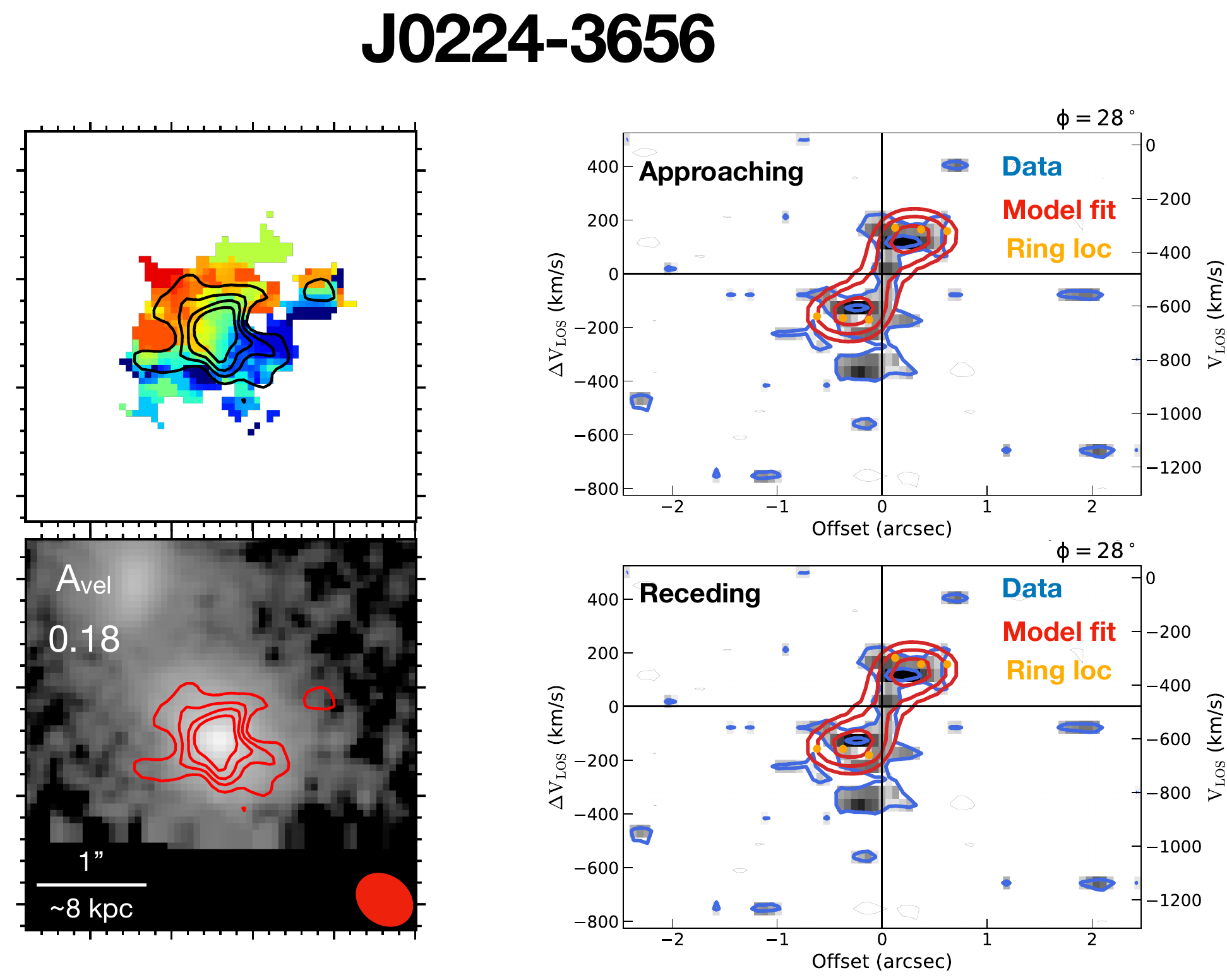}
	\caption{\textbf{Top Left:} Caption same as Figure \ref{fig:appendix_pvd_asym}.}
	\label{fig:additional}
\end{figure*}

\begin{figure*}
    \epsscale{0.85}
	\plotone{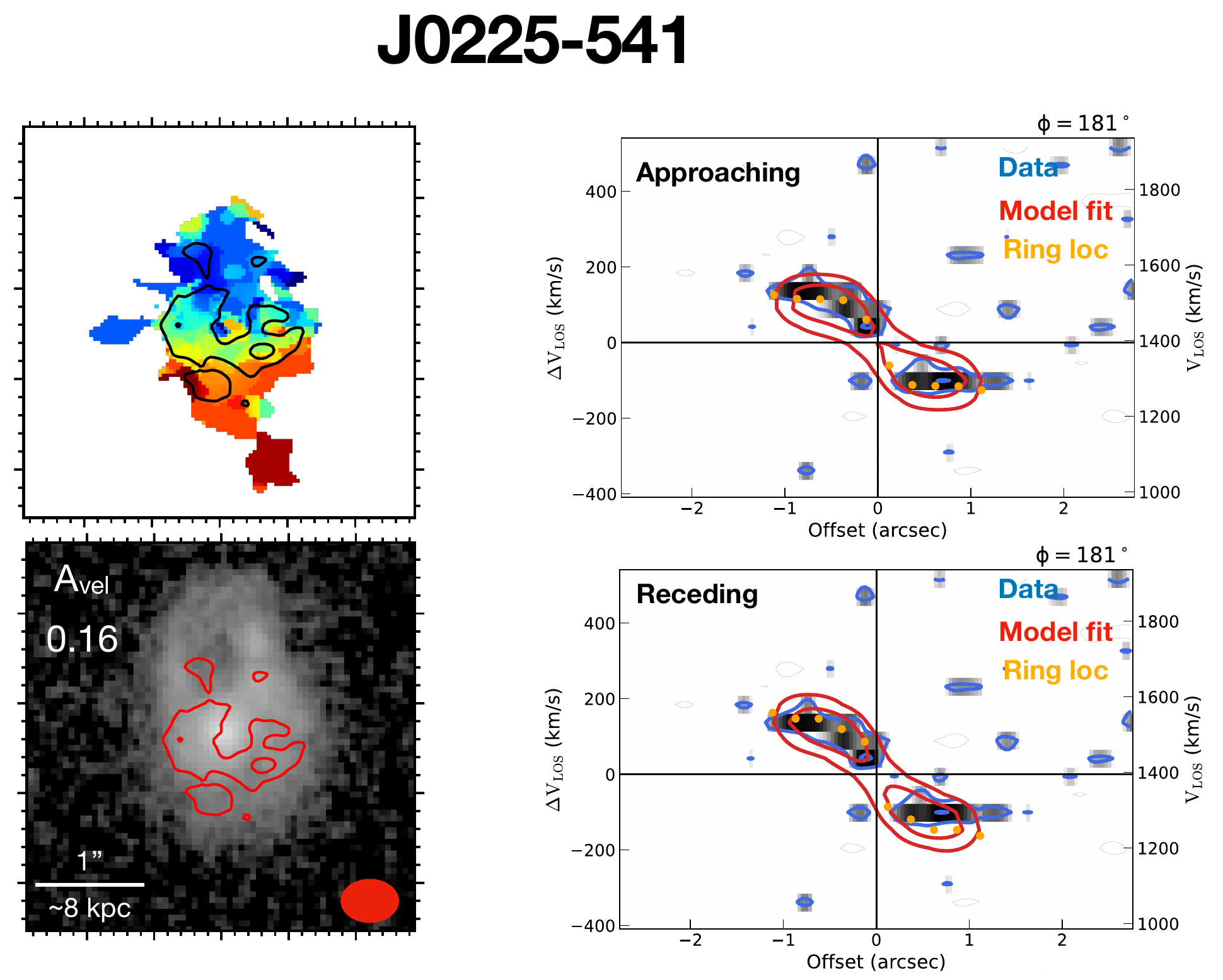}
	\caption{\textbf{Top Left:} Caption same as Figure \ref{fig:appendix_pvd_asym}.}
	\label{fig:additional}
\end{figure*}

\begin{figure*}
    \epsscale{0.85}
	\plotone{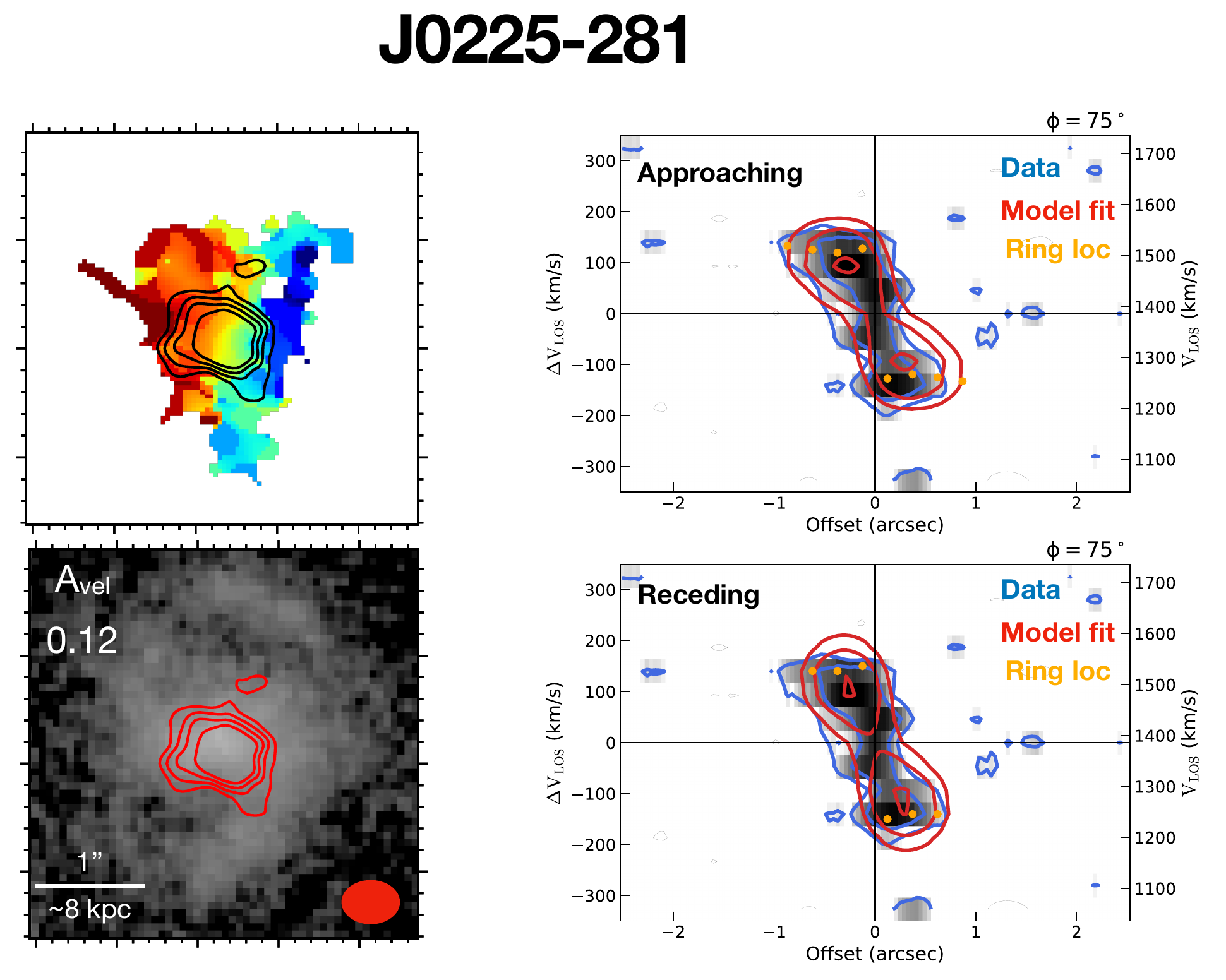}
	\caption{\textbf{Top Left:} Caption same as Figure \ref{fig:appendix_pvd_asym}.}
	\label{fig:additional}
\end{figure*}

\begin{figure*}
    \epsscale{0.85}
	\plotone{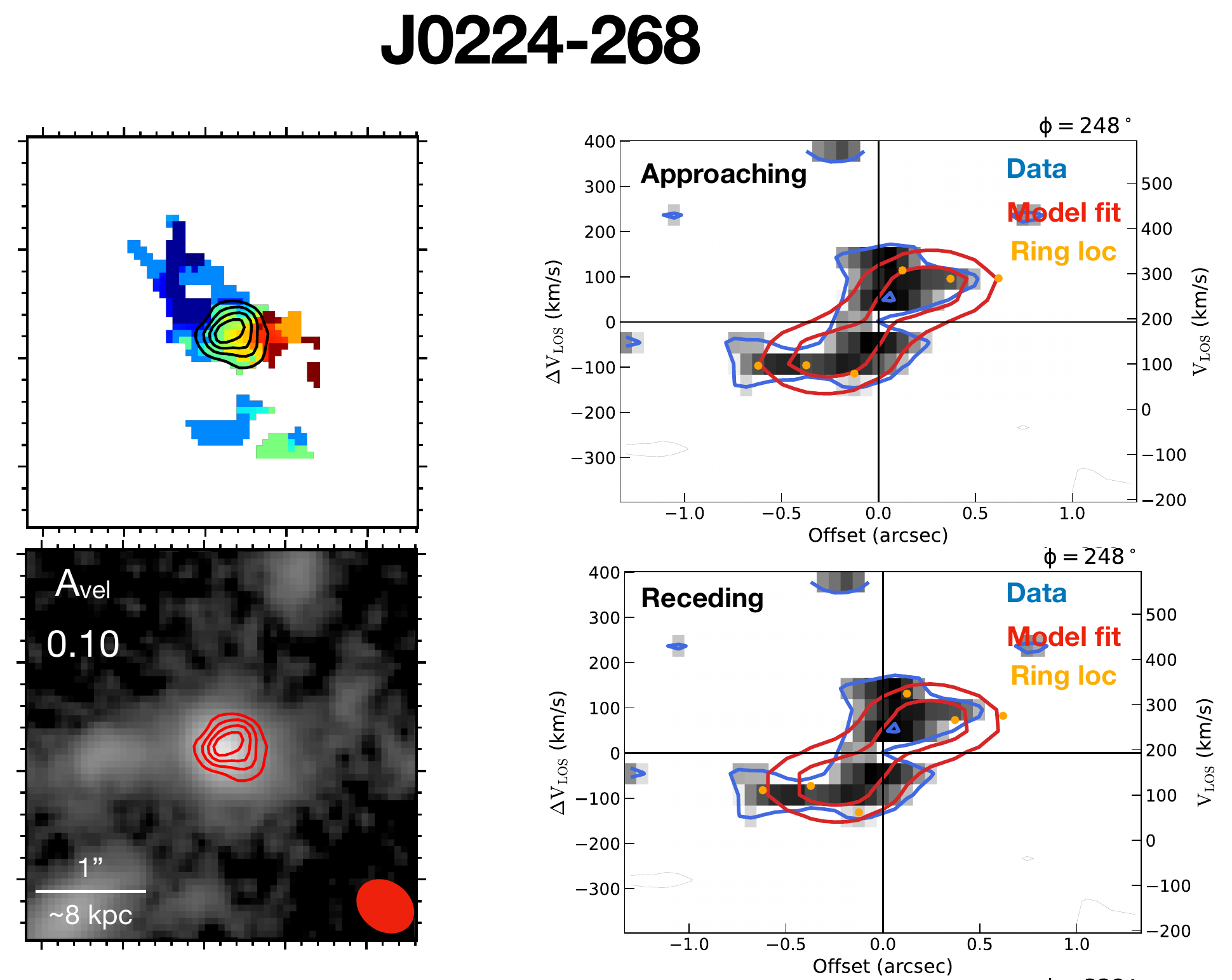}
	\caption{\textbf{Top Left:} Caption same as Figure \ref{fig:appendix_pvd_asym}.}
	\label{fig:additional}
\end{figure*}

\begin{figure*}
    \epsscale{0.85}
	\plotone{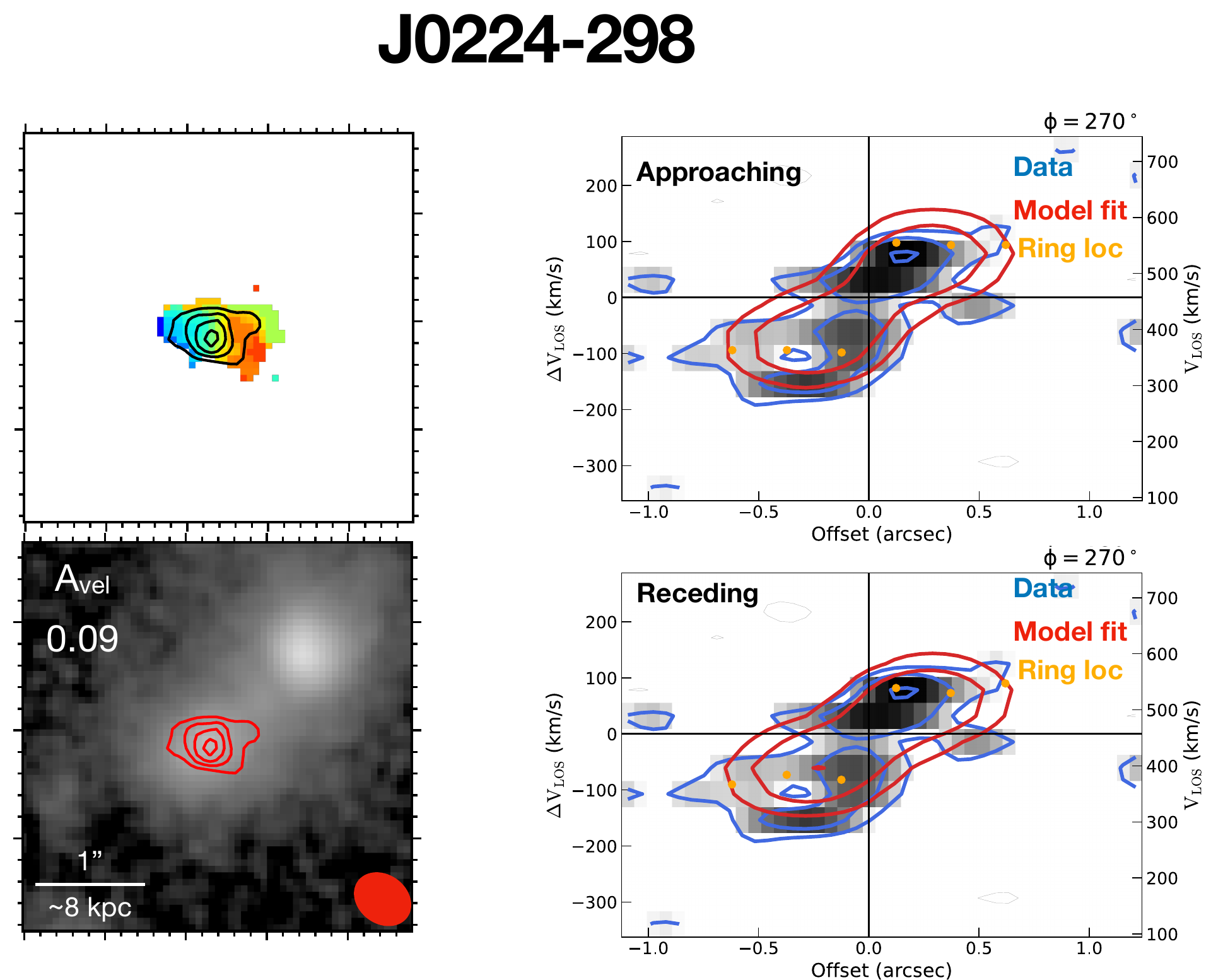}
	\caption{\textbf{Top Left:} Caption same as Figure \ref{fig:appendix_pvd_asym}.}
	\label{fig:additional}
\end{figure*}

\begin{figure*}
    \epsscale{0.85}
	\plotone{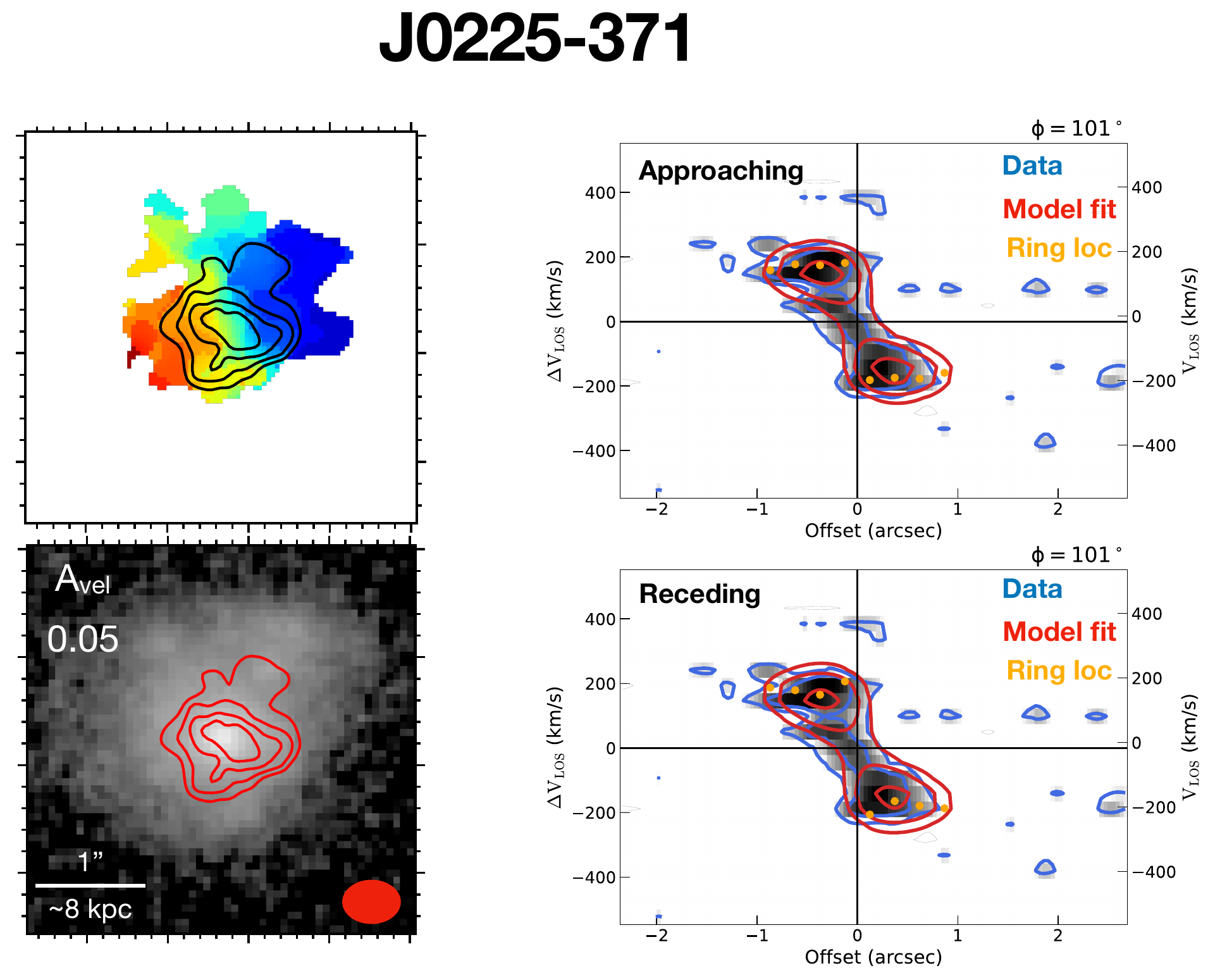}
	\caption{\textbf{Top Left:} Caption same as Figure \ref{fig:appendix_pvd_asym}.}
	\label{fig:additional}
\end{figure*}

\begin{figure*}
    \epsscale{0.85}
	\plotone{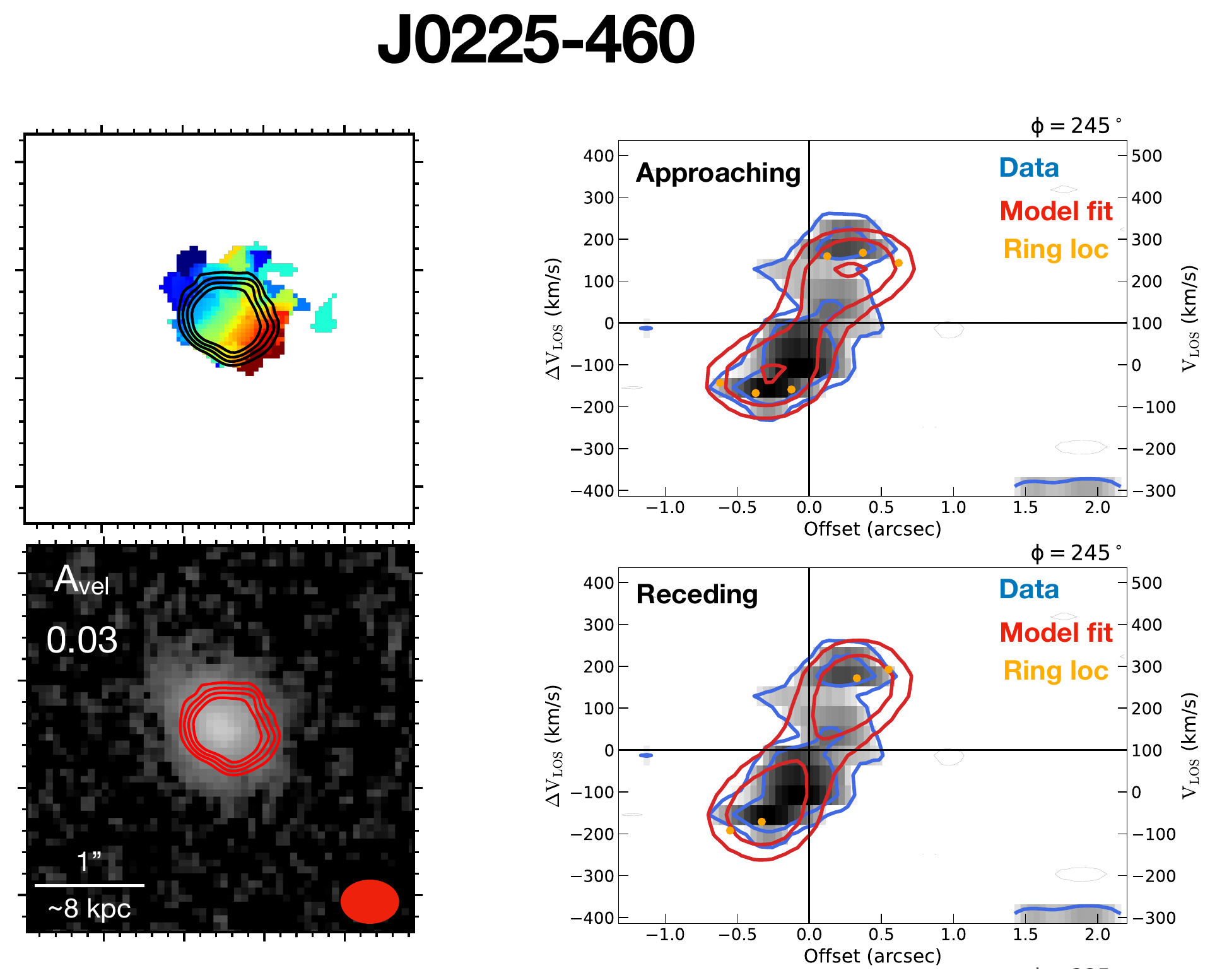}
	\caption{\textbf{Top Left:} Caption same as Figure \ref{fig:appendix_pvd_asym}.}
	\label{fig:additional}
\end{figure*}

\begin{figure*}
    \epsscale{0.85}
	\plotone{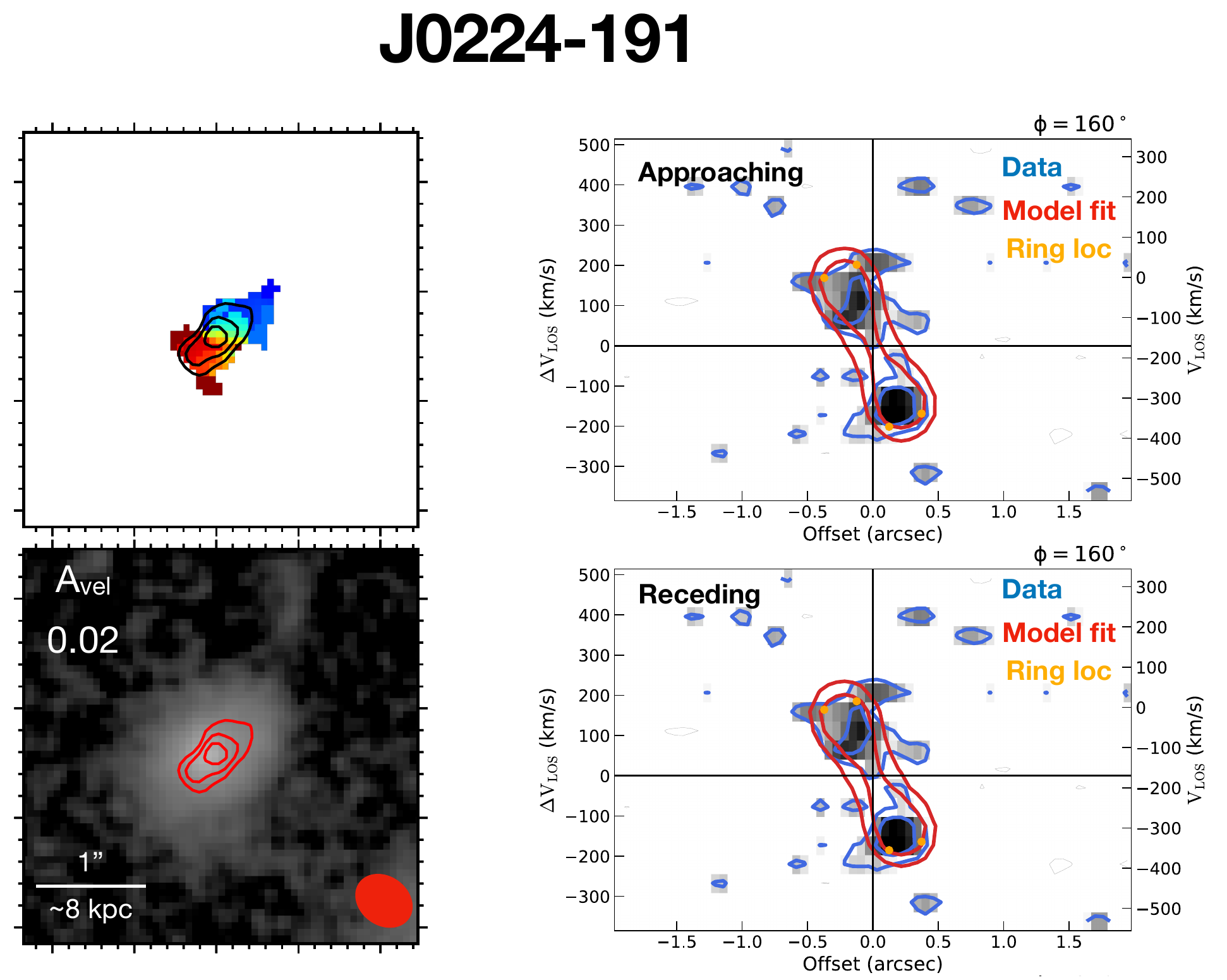}
	\caption{\textbf{Top Left:} Caption same as Figure \ref{fig:appendix_pvd_asym}.}
	\label{fig:additional}
\end{figure*}

\end{document}